\documentclass[lettersize,journal]{IEEEtran}
\usepackage{amsmath,amssymb,amsfonts}
\usepackage{array}
\usepackage[caption=false,font=normalsize,labelfont=sf,textfont=sf]{subfig}
\usepackage{textcomp}
\usepackage{stfloats}
\usepackage{url}
\usepackage{verbatim}
\usepackage{graphicx}
\usepackage{cite}
\usepackage{multirow}
\usepackage{booktabs}
\usepackage{xcolor}
\usepackage{makecell}
\usepackage{booktabs,dcolumn}
\usepackage[linesnumbered,ruled,vlined]{algorithm2e}
\usepackage{arydshln}
\newtheorem{prop}{Proposition}

    \def\Complex{{\rm\rule[.23ex]{.03em}{1.1ex}\kern-.3em{C}}}

    \newcommand{\be}{\begin{equation}} \newcommand{\ee}{\end{equation}}
    \newcommand{\bea}{\begin{eqnarray}} \newcommand{\eea}{\end{eqnarray}}
    \newcommand{\benum}{\begin{enumerate}} \newcommand{\eenum}{\end{enumerate}}

    \newcommand{\qa}{{\bf a}}
    \newcommand{\qb}{{\bf b}}

    \newcommand{\qh}{{\bf h}}

    \newcommand{\qo}{{\bf o}}
    \newcommand{\qp}{{\bf p}}
    \newcommand{\qq}{{\bf q}}

    \newcommand{\qu}{{\bf u}}
    \newcommand{\qv}{{\bf v}}
    \newcommand{\qw}{{\bf w}}
    \newcommand{\qx}{{\bf x}}

    \newcommand{\qG}{{\bf G}}
    \newcommand{\qH}{{\bf H}}
    \newcommand{\qI}{{\bf I}}
    \newcommand{\qJ}{{\bf J}}

    \newcommand{\qP}{{\bf P}}
    \newcommand{\qQ}{{\bf Q}}
    \newcommand{\qR}{{\bf R}}

    \newcommand{\qV}{{\bf V}}
    \newcommand{\qW}{{\bf W}}
    \newcommand{\qX}{{\bf X}}

    \newcommand{\qOmega}{{\boldsymbol \Omega}}

    \newcommand{\qalpha}{{\boldsymbol \alpha}}
    
    \newcommand{\qdelta}{{\boldsymbol \delta}}

    \newcommand{\qtau}{{\boldsymbol \tau}}

    \newcommand{\qvarepsilon}{{\boldsymbol \varepsilon}}

    \newcommand{\bbR}{{\mathbb R}}

    \newcommand{\calA}{{\cal A}}

    \newcommand{\calD}{{\cal D}}
    
    \newcommand{\calF}{{\cal F}}

    \newcommand{\tr}{{\sf tr}}

        \newcommand*{\argmin}{\operatornamewithlimits{argmin}\limits}
        \newcommand*{\argmax}{\operatornamewithlimits{argmax}\limits}

\begin{document}

\title{Machine Learning-Based Direct Source Localization for Passive Movement-Driven Virtual Large Array}

\author{
~Shang-Ling~Shih,~Chao-Kai~Wen,~\IEEEmembership{Fellow,~IEEE},~Chau~Yuen,~\IEEEmembership{Fellow,~IEEE}, and~Shi~Jin,~\IEEEmembership{Fellow,~IEEE}
\thanks{{S.-L.~Shih} and {C.-K.~Wen} are with the Institute of Communications Engineering, National Sun Yat-sen University, Kaohsiung 80424, Taiwan, Email: {\rm monlylonly@gmail.com, chaokai.wen@mail.nsysu.edu.tw}.}
\thanks{{C.~Yuen} is with the School of Electrical and Electronics Engineering, Nanyang Technological University, 639798 Singapore, Email: {\rm chau.yuen@ntu.edu.sg}.}
\thanks{{S.~Jin} is with the National Mobile Communications Research Laboratory, Southeast University, Nanjing 210096, P. R. China, Email: {\rm jinshi@seu.edu.cn}.}
}

\maketitle

\begin{abstract}

This paper introduces a novel smartphone-enabled localization technology for ambient Internet of Things (IoT) devices, leveraging the widespread use of smartphones. By utilizing the passive movement of a smartphone, we create a virtual large array that enables direct localization using only angle-of-arrival (AoA) information. Unlike traditional two-step localization methods, direct localization is unaffected by AoA estimation errors in the initial step, which are often caused by multipath channels and noise. However, direct localization methods typically require prior environmental knowledge to define the search space, with calculation time increasing as the search space expands. To address limitations in current direct localization methods, we propose a machine learning (ML)-based direct localization technique. This technique combines ML with an adaptive matching pursuit procedure, dynamically generating search spaces for precise source localization. The adaptive matching pursuit minimizes location errors despite potential accuracy fluctuations in ML across various training and testing environments. Additionally, by estimating the reflection source's location, we reduce the effects of multipath channels, enhancing localization accuracy. Extensive three-dimensional ray-tracing simulations demonstrate that our proposed method outperforms current state-of-the-art direct localization techniques in computational efficiency and operates independently of prior environmental knowledge.

\end{abstract}


\section{Introduction}

\IEEEPARstart{M}{utual} localization for Internet of Things (IoT) devices and user equipment (UE) is essential for various applications \cite{IoT}. Localization accuracy depends on precise distance and angle measurements. Common techniques for distance estimation include received signal strength (RSS), time-of-arrival (ToA), time-difference-of-arrival (TDoA), and round-trip time-of-flight (RToF). However, obstacles and mode mismatches can distort distance estimates obtained through RSS, while the need for time synchronization in ToA and multi-device coordination in TDoA adds complexity to these approaches. Although RToF can assist with time synchronization, it is not universally supported and is affected by transmission delays \cite{RToF1, RToF2}. These limitations have led to an alternative approach with angle-of-arrival (AoA)-based localization, which enables passive source localization without requiring synchronized timing.

Recent advancements in communication systems have enabled the equipping of multiple antennas on UEs, such as smartphones and wearables, facilitating AoA measurements. Large arrays can detect near-field signals, which carry both distance and angle information, enabling the localization of signal sources from a single site \cite{near_field_1, near_field_2}.\footnote{With a small array, only a single AoA from the target is obtained, making precise location estimation impossible. In contrast, a large array in the near-field zone, where propagation behaves as a spherical wave, captures multiple AoAs, allowing for accurate localization by using both distance and angular information from a single site.} However, incorporating such large arrays into UEs poses challenges, as large arrays are typically installed on base stations. Fortunately, as the UE moves---whether through natural motions like walking or by manually swaying the device---it can aggregate data from different positions, creating a virtual large array (VLA). This VLA enables high-accuracy localization of IoT devices in its near-field area. In this paper, we distinguish this VLA from a real large array (RLA), where all antennas are mounted on the same ground plane.

Traditionally, large arrays employ two-step localization methods, incorporating AoA estimation and object localization. By extracting parameters embedded in multipath components, studies like \cite{2step_ChannelSLAM, 2step_ROVER, 2step_AMA, 2step_EasyAPPos, 2step_MetaSLAM} introduced a simultaneous localization and mapping (SLAM) framework for localizing signal sources or reflectors. \cite{2step_BPLE} proposed a closed-form Bayesian solution to address the complex relationship between AoA estimation and source localization. Since AoA estimation and source localization have been deeply explored within their domains, the two-step approaches execute independent processes, easing integration. However, the AoA estimation errors caused by the multipath effect and noise in the first step can adversely affect localization in the second step, making direct methods more desirable.

Numerous studies have explored direct localization methods utilizing near-field signals with large arrays. For example, \cite{SOTA_SLAM} extended two-step SLAM to direct SLAM, avoiding any preprocessing stage that results in the loss of localization information. \cite{SOTA_MVDR} utilized distributed receivers as a VLA, employing a minimum-variance-distortionless-response for direct localization. \cite{SOTA_RIS_NLoS, SOTA_par_decouple, SOTA_RIS_LoS} leveraged reflected near-field signals from reconfigurable intelligent surfaces with RLAs for UE localization via maximum likelihood estimation. \cite{SOTA_NM} localized the target by applying a maximum likelihood estimator to the RLA, with and without exploiting carrier phase information. Meanwhile, \cite{SOTA_Doppler} focused on the Doppler effect in fast-moving arrays for enhanced localization accuracy. Multiple signal classification (MUSIC) algorithms were applied to RLAs and VLAs for direct localization in \cite{SOTA_ImperfectRLA,SOTA_MUSIC,SOTA_pathplan}. Additionally, \cite{SOTA_ELAA1} and \cite{SOTA_ELAA2} introduced a Newtonized orthogonal matching pursuit (NOMP) approach \cite{NOMP}, a super-resolution technique, for RLAs. These direct localization methods demonstrated localization accuracy that surpasses two-step methods.

Most existing direct localization methods have utilized grid search methods and relied on pre-existing environmental data to define the search space, thus becoming impractical for the user due to the dependency on measuring room size and the computational complexity that increases with room size. Additionally, \cite{SOTA_ELAA1} and \cite{SOTA_ELAA2} enhanced accuracy by estimating virtual sources (VSs) alongside physical sources (PSs), but obstacles and furniture create difficulties in determining VSs' search space due to variable reflections. To streamline these problems, \cite{SOTA_par_decouple} proposed an angle-decoupled grid search method suitable only for uniform planar RLAs, with the challenge of range definition remaining. Furthermore, \cite{SOTA_ELAA2} utilized array aperture to define the RLA's search space. However, noting that as UE moves, the VLA's array aperture also increases, paradoxically expanding the search space and computational demand. These practical challenges and their impact on dynamic UE localization efficiency were underexplored in the literature.

Some studies have explored machine learning (ML) for localization. Although ML requires extensive data and time at the offline training stage, it enables quick predictions during the online stage. Inspired by fingerprinting localization methods for static UEs \cite{fingerprint1} that can determine the UE's location in line-of-sight (LoS) or non-LoS (NLoS) scenarios, research such as \cite{ML_par_extract, ML_PNN, ML_spectrogram, ML_angle_delay, ML_all_CSI} fed location-bearing parameters or channel state information into the ML model. Moreover, fingerprinting localization methods can also be applied to large arrays \cite{fingerprint2}. Studies such as \cite{ML_2step} and \cite{ML_AAResCNN} achieved VLA localization methods by integrating ML with a two-step approach. \cite{ML_RLA, ML_RLA2, ML_RLA3} accomplished RLA localization through ML in a single step by feeding the covariance matrix of near-field signals into the ML model to locate the target. These studies have shown exceptional performance in online stage localization; however, the effectiveness of ML diminishes when the training and test data come from different indoor environments.

In response to these challenges, we introduce a feasible direct localization method for VLAs formed by the UE's passive movement. Our main contributions are as follows:
\begin{itemize}
\item \emph{Performance analysis for VLA:}
    Unlike RLAs, where all antennas synchronize to a single timing clock, VLAs treat antennas at disparate locations as components of a unified array, leading to variable time offsets as signals are received at different times. This distinction creates a localization performance gap between RLAs and VLAs. By utilizing the Cram'{e}r-Rao Lower Bound (CRLB), we analyze this performance gap and identify scenarios where VLAs can achieve performance on par with RLAs, considering the impracticality of equipping UEs with RLAs.

\item \emph{Efficient direct localization method on OFDM systems:}
    We develop a direct localization algorithm based on an orthogonal frequency-division multiplexing (OFDM) system, aligning with current broadband wireless systems. To address the challenges of defining the search space, we introduce ML-NOMP, a novel direct localization method that leverages ML to initialize the search space for PS within the NOMP framework. This approach enables rapid prediction of the PS's location without requiring prior environmental information.\footnote{This concept has been preliminarily verified in the conference version of this study \cite{DL_NOMP}.} Although predictions may be less accurate in unfamiliar indoor environments, Newton's refinement within the NOMP framework significantly reduces the margin of error. Once the PS's location is estimated, the propagation distances of the VSs can be determined, helping define the search space for the VSs. Estimating the VS locations can reduce the impact of multipath and enhance PS localization accuracy.

\item \emph{Simulation analysis in diverse environments:}
    Our simulations, conducted in extensive three-dimensional (3D) indoor environments, demonstrate that ML-NOMP not only outperforms existing state-of-the-art (SOTA) direct localization methods but also significantly reduces computational time, eliminates the need for prior environmental data, and delivers robust performance across diverse indoor settings, confirming its practical applicability.

\end{itemize}

\emph{Notations}---In this paper, variables, vectors, and matrices are denoted by italic letters $x$, small bold letters $\qx$, and capital bold letters $\qX$, respectively. The absolute value, phase, and real part of a variable $x$ are represented by $|x|$, $\angle{x}$, and $\Re\{x\}$, respectively. The Euclidean norm of the vector $\qx$ is denoted by $\|\qx\|$. $\qX^{\mathrm{T}}$, $\qX^{\mathrm{H}}$, and $\qX^{-1}$ signify the transpose, Hermitian transpose, and inverse of the matrix $\qX$, respectively. The $m$-th element of the vector $\qx$ is indicated by $[\qx]_m$, and the element at the $m$-th row and $n$-th column of the matrix $\qX$ is denoted by $[\qX]_{m,n}$. The determinant and trace of matrix $\qX$ is represented by ${\sf det}(\qX)$ and $\tr(\qX)$, respectively. $\qI_M$ is an $M \times M$ identity matrix. The Kronecker product is represented by $\otimes$.

\begin{figure}
\centering
\includegraphics[width=3.2in]{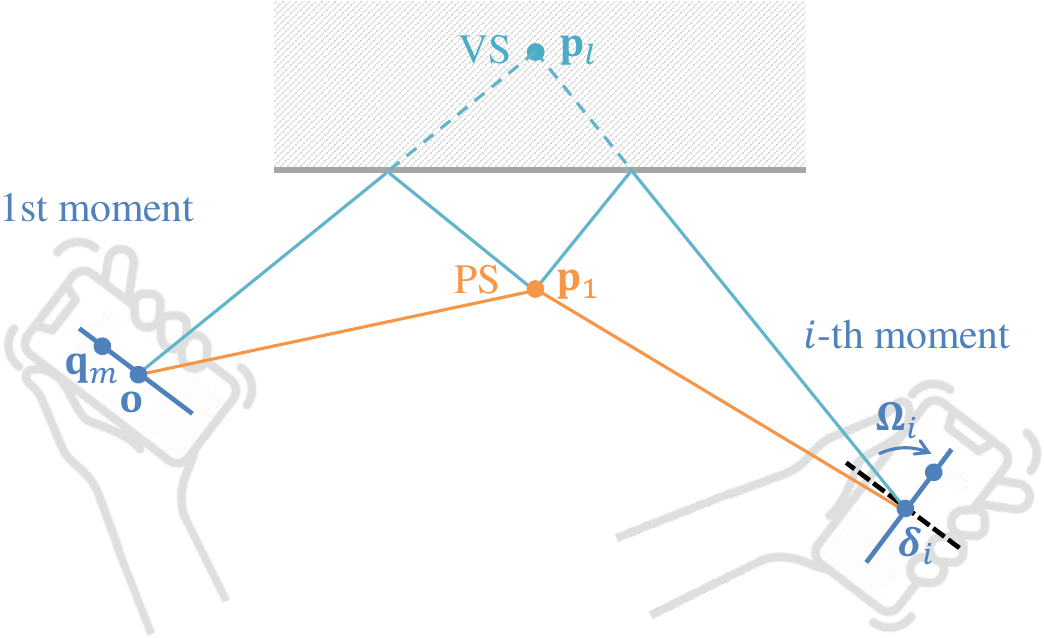}
\caption{Locations of the PS, VS, and the moving UE.}
\label{fig:IoT&UE}
\end{figure}

\section{System Model}
\label{sec:System_Model}

This section presents the signal model for VLAs, shaped by array movement and rotation, in array signal processing. We start with the coordinate system foundation in Section \ref{subsec:Coordinate_System}, detail antenna responses pre- and post-movement and rotation in Sections \ref{subsec:Before_M&R} and \ref{subsec:After_M&R}, and conclude with the VLA target signal model in Section \ref{subsec:Model_VLA}.

\subsection{Coordinate System}
\label{subsec:Coordinate_System}

In the 3D coordinate system, a vector pointing from location $\qv_1=[v_1^{\rm x}, v_1^{\rm y}, v_1^{\rm z}]^{\rm T}$ to location $\qv_2=[v_2^{\rm x}, v_2^{\rm y}, v_2^{\rm z}]^{\rm T}$ is expressed as $\qv = \qv_2 - \qv_1 = [v^{\rm x}, v^{\rm y}, v^{\rm z}]^{\rm T}$ in the Cartesian coordinates and can be transformed into spherical coordinates as
    \begin{equation}
    \label{eq:spherical_model}
    \qv = d \qu(\phi, \theta),
    \end{equation}
where
    \begin{equation*}
    \begin{aligned}
    \label{eq:spherical_distance}
    d &= \sqrt{(v^{\rm x})^2 + (v^{\rm y})^2 + (v^{\rm z})^2}, \\
    \phi =  &\arcsin {\left(  \frac{v^{\rm z}}{d}  \right)}, ~~ \theta = \arctan {\left(  \frac{v^{\rm y}}{v^{\rm x}}  \right)}
    \end{aligned}
    \end{equation*}
denote the length, elevation, and azimuth of the vector $\qv$, respectively, and
    \begin{equation*}
    \label{eq:direction_vector}
    \qu(\phi, \theta) = [\cos(\phi)\cos(\theta),\, \cos(\phi)\sin(\theta),\, \sin(\phi)]^{\rm T}
    \end{equation*}
denotes a unit vector with direction $(\phi, \theta)$.

\subsection{Array Response Before Moving and Rotation}
\label{subsec:Before_M&R}

We assume that the IoT device and the UE are equipped with one omnidirectional antenna and a patch antenna array, respectively. Initially, let us consider the global coordinate system. We assume the UE is located at the origin $\qo$ at the first moment and the IoT device at $\qp_1 = [p_1^{\rm x}, p_1^{\rm y}, p_1^{\rm z}]^{\rm T}$, as illustrated in Fig. \ref{fig:IoT&UE}.\footnote{If the UE is not at the origin, the locations of all other entities can be adjusted by subtracting the UE's location. Consequently, we can shift the UE's position to the origin, ensuring that all arguments presented in this paper remain valid.} The signals received by the UE originate from the IoT device through $L$ propagation paths, including the LoS and reflected paths. The reflected paths are considered to be generated from VSs located at $\qp_2,\ldots,\qp_L$.
Analogous to \eqref{eq:spherical_model}, the vector pointing from the UE's center to the $l$-th source is written as
    \begin{equation}
    \label{eq:UE_to_source_v1}
    \qp_l = d_l \qu(\phi_l, \theta_l),
    \end{equation}
where $d_l$, $\phi_l$, and $\theta_l$ represent the distance, elevation, and azimuth of the $l$-th source to UE's center, respectively.

The UE is equipped with an array of $M$ antennas. Transitioning to the local coordinate system from the UE's perspective, the vector pointing from the UE's center to its $m$-th antenna, corresponding to the antenna's location, is written as
    \begin{equation}
    \label{eq:UE_to_antenna_v1}
    \qq_m =  r_m \qu(\psi_m, \vartheta_m),
    \end{equation}
where $r_m$, $\psi_m$, and $\vartheta_m$ indicate the distance, elevation, and azimuth of UE's $m$-th antenna to its center, respectively. From \eqref{eq:UE_to_source_v1} and \eqref{eq:UE_to_antenna_v1}, the vector from the UE's $m$-th antenna to the $l$-th source is
    \begin{equation}
    \label{eq:antenna_to source_v1}
    \qp_l - \qq_m = d_l \qu(\phi_l, \theta_l) - r_m \qu(\psi_m, \vartheta_m),
    \end{equation}
and the length of this vector is
    \begin{equation}
    \label{eq:antenna_to source_d1}
    \| \qp_l - \qq_m \|  = \sqrt{ {d_l^2} + {r_m^2} - 2{d_l}{r_m} {\qu^{\rm T}(\phi_l, \theta_l)} {\qu(\psi_m, \vartheta_m)} }.
    \end{equation}
If the source's distance is significantly greater than the array's size, i.e., $r_m \ll d_l$, \eqref{eq:antenna_to source_d1} simplifies as\footnote{The Taylor approximation at $x=0$ is $\sqrt{1+x} \approx 1 + {x}/{2} $. }
    \begin{align}
    \label{eq:far_field_d1}
     \| \qp_l - \qq_m \|  &\approx {d_l} - {r_m} {\qu^{\rm T}(\phi_l, \theta_l)} {\qu(\psi_m, \vartheta_m)} \notag \\
                                  &=          {d_l} -  {\qu^{\rm T}(\phi_l, \theta_l)} {\qq_m},
    \end{align}
following \eqref{eq:UE_to_antenna_v1}. This assumption is commonly used in the far-field scenario.

The signal propagation delays between the antennas and the UE's center induce phases in the received signals, forming the array response vector $\qa(\phi_l,\theta_l) \in \mathbb{C}^{M\times1}$. Specifically, the response at the $m$-th antenna due to the $l$-th source, which is the $m$-th element of the array response vector, is given by
    \begin{equation}
    \label{eq:array_response_1}
    {\left[\qa(\phi_l, \theta_l)\right]}_m =          e^{ -{\sf j} \frac{2 \pi}{\lambda} ( \| \qp_l - \qq_m \|-d_l) }
                                                         \approx e^{   {\sf j} \frac{2 \pi}{\lambda} {\qu^{\rm T}(\phi_l, \theta_l)}  {\qq_m}  },
    \end{equation}
where $\lambda$ denotes the wavelength, and the approximation is derived from \eqref{eq:far_field_d1}. Notably, the array response depends not on the source's distance, $d_l$, but solely on the direction, $(\phi_l, \theta_l)$. This model is often referred to as the far-field signal model, from which we can estimate the direction.

\subsection{Array Response After Moving and Rotation}
\label{subsec:After_M&R}

As depicted in Fig. \ref{fig:IoT&UE}, when a person naturally moves the UE while walking, the UE may shift to $\qdelta_{i}$ at the $i$-th moment. The vector from the UE's center to the $l$-th source is
    \begin{equation}
    \label{eq:UE_to_source_vi}
    \tilde{\qp}_{i,l}= \qp_l - \qdelta_i = \tilde{d}_{i,l} \qu(\tilde{\phi}_{i,l}, \tilde{\theta}_{i,l}),
    \end{equation}
where $\tilde{d}_{i,l}$, $\tilde{\phi}_{i,l}$, and $\tilde{\theta}_{i,l}$ denote the distance, elevation, and azimuth of the $l$-th source to UE's center at the $i$-th moment, respectively. In addition to movement, the shaking induces rotations on the UE. The orientation of the array within the UE can be succinctly described using a rotation matrix $\qOmega$. This matrix has the properties ${\qOmega^{\rm T}\qOmega = \qI_3}$ and ${\sf det}(\qOmega) = 1$. With $\qOmega_i$ representing the rotation at the $i$-th moment, the vector from the UE's center to its $m$-th antenna becomes
    \begin{equation}
    \label{eq:pm_i}
    \tilde{\qq}_{i,m} = \qOmega_i \qq_m = r_m \qOmega_i \qu(\psi_m, \vartheta_m),
    \end{equation}
following \eqref{eq:UE_to_antenna_v1}. Consequently, the vector from the UE's $m$-th antenna to the $l$-th source is
    \begin{equation}
    \label{eq:antenna_to source_vi}
    \tilde{\qp}_{i,l} - \tilde{\qq}_{i,m} = \tilde{d}_{i,l} \qu(\tilde{\phi}_{i,l}, \tilde{\theta}_{i,l}) - r_m \qOmega_i \qu(\psi_m, \vartheta_m),
    \end{equation}
and its approximate length, using the simplification in \eqref{eq:far_field_d1}, is
    \begin{equation}
    \label{eq:antenna_to source_di}
    \| \tilde{\qp}_{i,l} - \tilde{\qq}_{i,m} \| \approx \tilde{d}_{i,l} - \qu^{\rm T}(\tilde{\phi}_{i,l}, \tilde{\theta}_{i,l}) \qOmega_i \qq_m.
    \end{equation}
Thus, the response of the $m$-th antenna to the $l$-th source is
    \begin{equation}
    \label{eq:array_response_i}
    [\qa(\tilde{\phi}_{i,l}, \tilde{\theta}_{i,l})]_m \approx e^{ {\sf j} \frac{2 \pi}{\lambda} \qu^{\rm T}(\tilde{\phi}_{i,l}, \tilde{\theta}_{i,l}) \qOmega_i\qq_m }.
    \end{equation}

Given that the UE is equipped with internal sensors, the movement and rotation of the UE is readily available. Therefore, by applying \eqref{eq:UE_to_source_vi}, we can transform \eqref{eq:array_response_1} and \eqref{eq:array_response_i} into a generalized model for any given moment:
    \begin{equation}
    \label{eq:array_response_general}
    {\left[\qa(\qp_l;\, \qdelta_i, \qOmega_i)\right]}_m = e^{ {\sf j} \frac{2 \pi}{\lambda} \frac{ ( \qp_l- \qdelta_i )^{\rm T}}{\| \qp_l- \qdelta_i \|} \qOmega_i \qq_m }.
    \end{equation}
This model facilitates estimating the sources' location by incorporating the internal sensor information. Note that we assume the relative movement and rotation are obtained from other sensor systems; thus, they are not the unknown variables of interest in this context. We use the notation ``;" to distinguish these known variables from the other unknown parameters.

\subsection{Signal Model for the VLA}
\label{subsec:Model_VLA}

The IoT device operates an OFDM system, with reference signals spanning $N$ subcarriers. To pinpoint the location of the IoT device, we move the UE for a certain duration, capturing the channel frequency response (CFR) from the incoming signal during this period. At the $i$-th moment, the aggregated CFR across all antennas and subcarriers is expressed as
    \begin{equation}
    \label{eq:h}
    \qh_{i} = \sum_{l=1}^{L} { \alpha_{i,l} \qv(\qp_l, \tau_i;\, \qdelta_i, \qOmega_i) }  + \qw_i.
    \end{equation}
Here, $\alpha_{i,l}$ represents the complex-valued channel coefficient of the $l$-th path, $\tau_i$ is the time offset, and $\qw_{i}\sim\mathcal{CN}(\boldsymbol{0}_{NM},\sigma^2\qI_{NM})$ denotes the white Gaussian noise vector. We define
    \begin{equation}
    \label{eq:v}
    \qv(\qp_l, \tau_i;\, \qdelta_{i}, \qOmega_{i}) = \qa(\qp_l;\, \qdelta_i, \qOmega_i) \otimes \qb(\qp_l, \tau_i;\, \qdelta_i),
    \end{equation}
where $\qa(\qp_l;\, \qdelta_i, \qOmega_i) \in \mathbb{C}^{M\times1}$ is the aggregate of the $M$ antenna responses, as outlined in \eqref{eq:array_response_general}. Meanwhile, $\qb(\qp_l, \tau_i;\, \qdelta_i)\in \mathbb{C}^{N\times1}$ captures the time delay's frequency responses across $N$ subcarriers. In the OFDM system under consideration, the signal's propagation delays, along with time offsets, manifest as phases across each subcarrier. Thus, the $n$-th element of $\qb(\qp_l, \tau_i;\, \qdelta_i)$ is given by
    \begin{equation}
    \label{eq:subcarrier_response_general}
    {\left[\qb(\qp_l, \tau_i;\, \qdelta_i)\right]}_n =  e^{-{\sf j} 2 \pi  f_n {\left( \frac{\|\qp_l-\qdelta_i\|}{c} - {\tau_i} \right)} },
    \end{equation}
where $f_n$ is the relative frequency of the $n$-th subcarrier against the carrier frequency, and $c$ denotes the speed of light. We define the subcarrier frequency as $[f_1,...,f_N]= [-N/2, \dots , -1, 1, \dots, N/2] \cdot \Delta f$, where $\Delta f$ is the subcarrier frequency spacing.

Notably, in the VLA system, the time offsets $\tau_i$ for each $i$ vary because we exclude using a dedicated transmitter for time synchronization. This variation makes it challenging to determine the ToAs from a single CFR snapshot, making \emph{direct localization} from AoA measurements alone difficult. However, by aggregating the CFR with measured movements and rotations of the UE across $I$ instances, a near-field signal model is constructed even without using RLA. This setup enables direct localization using the set $\{ \qh_{1}, \ldots, \qh_{I} \}$.

\section{Performance Bounds}

This section explores the performance bounds of direct localization using VLAs, laying the foundation for positioning with passively moved UEs. We calculate the CRLB for localizing PS via VLA's CFR. Given the lack of geometric information about the indoor environment, the sources' positions are estimated independently. Therefore, in the subsequent subsections, we focus on the LoS channel for CRLB analysis as an illustrative example. This analysis can be readily extended to multipath channels by considering propagation paths from VSs.

To assess effectiveness, we also compare the capabilities of the VLA with those of an RLA. The RLA consists of several modules with array identical to the UE's array, each aligned with the stance of the UE at successive moments. The primary difference between the RLA and VLA lies in the CFR capture process: RLA modules capture the CFR simultaneously, resulting in the same time offsets $\tau$ across all modules, while the VLA captures CFR continuously as it moves, leading to varying time offsets $\tau_i$ at each moment. Although the RLA is expected to outperform the VLA in terms of IoT device localization, equipping an RLA in a UE is impractical due to the UE's typically small form factor. Our CRLB analysis proposes scenarios where VLAs can achieve localization accuracy comparable to RLAs, addressing the practical challenges of using RLAs.

Notably, a related CRLB analysis by \cite{SOTA_NM} compared RLA localization performance with and without utilizing carrier phase information, assuming coherent RLA modules with a uniform phase offset. This approach allowed the extraction of distance-related data from the phase of the complex channel coefficient $\alpha_{i,l}$. Unlike \cite{SOTA_NM}, our study focuses on non-coherent RLA and VLA, deliberately excluding distance information from $\alpha_{i,l}$ as carrier phase exploitation is beyond our scope. Consequently, we avoid configuring RLA modules in coherent mode, aligning our RLA setup more closely with the VLA.

\subsection{FIM of VLA}

For ease of notation, we remove the path index $l$ from \eqref{eq:h} and present the CFR in matrix form:
    \begin{equation}
    \label{eq:VLA_channel}
    \qH_{i} = \alpha_i \qb(\qp, \tau_i;\, \qdelta_i) \qa(\qp;\, \qdelta_i, \qOmega_i) ^{\rm T}  + \qW_i,
    \end{equation}
where $\qp$ and $\alpha_i$ come from $\qp_1$ and $\alpha_{i,1}$, respectively.\footnote{In this Section, we use $\qp$ or $\qp_1$ interchangeably.} We introduce the vector that collects all the unknown parameters that need to be estimated:
    \begin{equation}
    \label{eq:VLA_parameters}
    \qvarepsilon =  \left[ \qp^{\rm T}, \tau_1, \ldots, \tau_I, \gamma_1, \beta_1, \ldots, \gamma_I, \beta_I  \right]^{\rm T} \in \bbR^{3+3I},
    \end{equation}
where ${\gamma_i = \left| \alpha_i \right|}$ and ${\beta_i = \angle \alpha_i}$. The first three elements on $\qvarepsilon$ are related to the localization, and the remaining elements are related to the channel propagation. The CRLB is the inverse of the Fisher information matrix (FIM). The element of FIM on the $s$-th row and $s'$-th column is defined as
    \begin{equation}
    \label{eq:VLA_total_FIM}
    \left[\qJ \right]_{s,s'} = \qJ^{\varepsilon_s, \varepsilon_{s'}} = \sum_{i=1}^{I} \qJ_{i}^{\varepsilon_s, \varepsilon_{s'}},
    \end{equation}
where
    \begin{equation}
    \label{eq:VLA_single_FIM}
    \qJ_{i}^{\varepsilon_s, \varepsilon_{s'}} = \frac{2}{\sigma^2} \Re \left\{ \sum_{n=1}^{N} \sum_{m=1}^{M}
                                                                                                                     \frac{\partial \left[ \qH_i \right]_{m,n}^{*}}{\partial [\qvarepsilon]_s}
                                                                                                                     \frac{\partial \left[ \qH_i \right]_{m,n}}{\partial [\qvarepsilon]_{s'}} \right\}.
    \end{equation}

In \eqref{eq:VLA_single_FIM}, the partial derivatives of the CFR with respect to all the parameters in $\qvarepsilon$ are presented in \eqref{eq:VH}, shown in Appendix A. By substituting \eqref{eq:VH} into \eqref{eq:VLA_single_FIM}, we can further derive the equations used to calculate the upper triangular portions of the FIM, shown in \eqref{eq:VJ_i} in Appendix A. Since the FIM is a symmetric matrix, the lower triangular portions mirror those above the diagonal. By substituting \eqref{eq:VJ_i} into \eqref{eq:VLA_total_FIM} and applying the Schur complement to the FIM $\qJ$, we obtain the equivalent FIM (EFIM) for localization as
    \begin{equation}
    \label{eq:VLA_SC}
    \qV^{\qp,\qp} = \sum_{i=1}^{I} {\left( \qJ_i^{\qp,\qp}   - \frac{\qJ_i^{\qp,\beta_i}     \qJ_i^{\beta_i,\qp}}    {\qJ_i^{\beta_i,\beta_i}} \right)}
                           -   \sum_{i=1}^{I} \frac{\qJ_i^{\qp,\tau_i} \qJ_i^{\tau_i, \qp}  }{\qJ_i^{\tau_i, \tau_i}}.
    \end{equation}

\subsection{FIM of RLA}

Similar to \eqref{eq:VLA_channel}, the LoS CFR of the RLA can be present as
    \begin{equation}
    \label{eq:RLA_channel}
    \bar{\qH}_{i} = \alpha_i \qb(\qp, \tau;\, \qdelta_i) \qa(\qp;\, \qdelta_i, \qOmega_i) ^{\rm T}  + \qW_i,
    \end{equation}
and the vector with the unknown parameters in the CFR is
    \begin{equation}
    \label{eq:RLA_parameters}
    \bar{\qvarepsilon} =  \left[ \qp^{\rm T}, \tau, \gamma_1, \beta_1, \ldots, \gamma_I, \beta_I  \right]^{\rm T} \in \bbR^{4+2I}.
    \end{equation}
The element of the FIM of the RLA on the $s$-th row and $s'$-th column is defined as
    \begin{equation}
    \label{eq:RLA_total_FIM}
    \left[\bar{\qJ} \right]_{s,s'} = \bar{\qJ}^{\bar{\varepsilon}_s, \bar{\varepsilon}_{s'}} = \sum_{i=1}^{I} \bar{\qJ}_{i}^{\bar{\varepsilon}_s, \bar{\varepsilon}_{s'}},
    \end{equation}
where
    \begin{equation}
    \label{eq:RLA_single_FIM}
    \bar{\qJ}_{i}^{\bar{\varepsilon}_s, \bar{\varepsilon}_{s'}} = \frac{2}{\sigma^2} \Re \left\{ \sum_{n=1}^{N} \sum_{m=1}^{M}
                                                                                                                     \frac{\partial \left[ \bar{\qH}_i \right]_{m,n}^{*}}{\partial [\bar{\qvarepsilon}]_s}
                                                                                                                     \frac{\partial \left[ \bar{\qH}_i \right]_{m,n}}{\partial [\bar{\qvarepsilon}]_{s'}} \right\}.
    \end{equation}

In \eqref{eq:RLA_single_FIM}, the partial derivatives of the CFR with respect to all the parameters in $\bar{\qvarepsilon}$ are presented in \eqref{eq:RH}. Substituting \eqref{eq:RH} into \eqref{eq:RLA_single_FIM} can further obtain \eqref{eq:RJ_i}. The results of \eqref{eq:RH} and \eqref{eq:RJ_i} are shown in Appendix B. Similar to \eqref{eq:VLA_SC}, by substituting  \eqref{eq:RJ_i} into \eqref{eq:RLA_total_FIM} and applying the Schur complement to the FIM $\bar{\qJ}$, we can obtain the EFIM for localization as
    \begin{multline}
    \label{eq:RLA_SC}
    \qR^{\qp,\qp}   = \sum_{i=1}^{I} {\left( \qJ_i^{\qp,\qp}   - \frac{\qJ_i^{\qp,\beta_i}  \qJ_i^{\beta_i,\qp}} {\qJ_i^{\beta_i,\beta_i}} \right)}
                            \\ -  \frac{ {\left( \sum_{i=1}^{I}  {\qJ}_i^{\qp,\tau_i} \right)} {\left( \sum_{i=1}^{I}  {\qJ}_i^{\tau_i, \qp} \right)} }{\sum_{i=1}^{I} {\qJ}_i^{\tau_i, \tau_i}}.
     \end{multline}

\subsection{Comparison of RLA and VLA}

Generally, localization performance is evaluated using the squared position error bound (SPEB), corresponding to the trace of the inverse EFIM. However, due to the complex nature of the $\qJ_i^{\qp,\qp}$ terms in the EFIMs given in \eqref{eq:VLA_SC} and \eqref{eq:RLA_SC}, inverting $\qV^{\qp,\qp}$ and $\qR^{\qp,\qp}$ poses significant challenges. Therefore, instead of the SPEB, we analyze the performance gap between RLA and VLA using relative results. Recall that the trace of a matrix equals the sum of its eigenvalues; thus, as the trace of the FIM increases, its eigenvalues increase, leading to a decrease in the eigenvalues of the CRLB and the SPEB. Consequently, we calculate the FIM trace rather than the SPEB, with the difference in FIM traces between RLA and VLA revealing factors that affect localization performance. By substituting \eqref{eq:VJ_i} into \eqref{eq:VLA_SC} and \eqref{eq:RLA_SC} and then subtracting \eqref{eq:VLA_SC} from \eqref{eq:RLA_SC}, we can derive the following proposition.

\begin{prop} \label{th:D_FIM}
The FIM difference between the RLA and the VLA is given by
    \begin{align}
    \label{eq:EFIM_diff_I}
    \Delta{\rm FIM}_I &\triangleq \qR^{\qp,\qp} - \qV^{\qp,\qp}, \notag\\
                                 =  &\sum_{i=1}^{I-1} \sum_{j=i+1}^{I} \zeta_{i,j}
                                            {\left( \frac{\tilde{\qp}_i}{\| \tilde{\qp}_i \|} - \frac{\tilde{\qp}_j}{\| \tilde{\qp}_j \|} \right)}
                                            {\left( \frac{\tilde{\qp}_i}{\| \tilde{\qp}_i \|} - \frac{\tilde{\qp}_j}{\| \tilde{\qp}_j \|} \right)}^{\rm T},
   \end{align}
where
    \begin{equation}
    \label{eq:zeta}
    \zeta_{i,j} = \frac{2 {\pi^2}  {\Delta f^2} M N(N+1)(N+2)}{3{\sigma^2}{c^2}} \cdot
    \frac{ {\gamma_i^2}  {\gamma_j^2}}{\sum_{k=1}^{I} \gamma_k^2},
    \end{equation}
and $\tilde\qp_i = \qp - \qdelta_i$. The trace of the $\Delta {\rm FIM}_I$ is given by
   \begin{equation}
   \label{eq:trace_diff_I}
   \Delta{\rm TR}_I \triangleq \tr( \Delta{\rm FIM}_I) = \sum_{i=1}^{I-1} \sum_{j=i+1}^{I} 2 \zeta_{i,j} (1 - \cos \rho_{i,j}),
   \end{equation}
where
    \begin{equation}
    \label{eq:rho}
    \rho_{i,j} = \arccos \left( \frac{\tilde{\qp}_i^{\rm T} \tilde{\qp}_j}{\| \tilde{\qp}_i \|\| \tilde{\qp}_j \|} \right)
    \end{equation}
is the angle between vectors $\tilde{\qp}_i$ and $\tilde{\qp}_j$.
\end{prop}

\begin{figure*}
\centering
\includegraphics[width=7in]{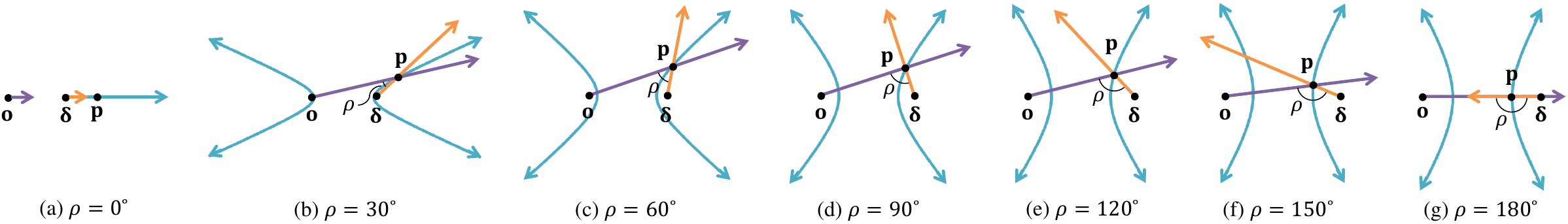}
\caption{The location candidates of $\qp$ determined by the angle information obtained at $\qo$ (purple line), the angle information obtained at $\qdelta$ (orange line), and the range difference information (blue line) for different values of $\rho$.}
\label{fig:CRLB}
\end{figure*}

From \eqref{eq:trace_diff_I}, we can deduce that $\Delta{\rm TR}_I \geq 0$, indicating that RLA provides more localization information than VLA. To gain further insights, we begin with the specific case where $I = 2$. For simplicity in this special case, we denote $\tilde{\qp}_1=\qp$ and $\tilde{\qp}_2=\qp - \qdelta$. Therefore, \eqref{eq:trace_diff_I} becomes
   \begin{equation}
   \label{eq:trace_diff_2}
   \Delta{\rm TR}_2  = 2 \zeta (1-\cos \rho),
   \end{equation}
where
    \begin{equation}
    \label{eq:zeta_12}
    \zeta = \frac{2 {\pi^2}  {\Delta f^2} M N(N+1)(N+2)}{3{\sigma^2}{c^2}} \cdot
    \frac{ {\gamma_1^2}  {\gamma_2^2}}{ {\gamma_1^2} + {\gamma_2^2} },
    \end{equation}
and $\rho$ is the angle between the vectors $\qp$ and $(\qp-\qdelta)$, as shown in Fig. \ref{fig:CRLB}. The factor $\zeta$ can be interpreted as a range information intensity (RII) \cite{NLN}, which depends on $\Delta f$, $M$, $N$, $\gamma_1$, $\gamma_2$, and $\sigma$, all of which are determined by the system and the propagation environment. The localization performance gap between RLA and VLA varies with the angle $\rho$.

To further understand the physical meaning of $\zeta$ and $\rho$, we consider the CFRs $\{\qh_1, \qh_2\}$ and $\{\bar{\qh}_1, \bar{\qh}_2\}$ captured by VLA and RLA, respectively. The antenna response in \eqref{eq:array_response_general} and the frequency response in \eqref{eq:subcarrier_response_general} encapsulate angular and range information. Direct localization involves an exhaustive search for the location $\qp$ that best matches these responses. From the antenna response perspective, both RLA and VLA are characterized by one unknown parameter, $\qp$, allowing two angular pieces of information to determine the location, provided the two angles are neither identical nor supplementary. However, from the frequency response perspective, VLA's frequency responses involve three unknown parameters, $(\qp, \tau_1, \tau_2)$, thereby excluding range information. Conversely, RLA's frequency responses involve only two unknown parameters, $(\qp, \tau)$, enabling the inference of the range difference between $\qp$ and $(\qp - \qdelta)$. Consequently, \eqref{eq:trace_diff_2} highlights RLA's localization performance gain due to this range difference information, as reflected in the terms $\zeta$ and $\rho$.

The factor $\zeta$ represents the effective RII, where the first term relates to the signal-to-noise ratio and the effective bandwidth of the system, and the second term is the harmonic mean of ${\gamma_1^2}$ and ${\gamma_2^2}$. The harmonic mean is used because the accuracy of the range difference depends on the delay estimation accuracy at both $\qp$ and $(\qp - \qdelta)$, and such accuracy is limited by the worse of the two. In the general case where $I > 2$, other $i$-th and $j$-th array module pairs can jointly provide additional gain, leading to the summation in \eqref{eq:trace_diff_I}.

To explain how $\Delta{\rm TR}_2$ in \eqref{eq:trace_diff_2} varies with $\rho$, we use $\calA$ and $\calD$ to represent the feasible solution sets of $\qp$ determined by the angle and range difference information, respectively. $\calA$ includes the location candidates determined at the origin $\qo$, denoted by $\calA_{\qo}$, and the location candidates determined at $\qdelta$, denoted by $\calA_{\qdelta}$. Fig. \ref{fig:CRLB} shows examples with fixed $\| \qp \|$ and $\| \qp - \qdelta \|$ for various $\rho$. The purple, orange, and blue lines represent $\calA_{\qo}$, $\calA_{\qdelta}$, and $\calD$, respectively.

As depicted in Fig. \ref{fig:CRLB}(a), $\calA_{\qo}$, $\calA_{\qdelta}$, and $\calD$ overlap along the same line when $\rho=0^{\circ}$. In this scenario, neither large array can estimate $\qp$ using  $\calA$, and $\calD$ provides no additional information for RLA, resulting in ${\Delta{\rm TR}_2=0}$. As $\rho$ increases, as shown in Figs. \ref{fig:CRLB}(b) and \ref{fig:CRLB}(c), the intersection angle between $\calA_{\qo}$ and $\calA_{\qdelta}$, which equals to $\rho$, becomes larger, allowing both large arrays to estimate $\qp$ more accurately. Additionally, $\calD$ forms a hyperbola and intersects $\calA$ at a wider angle as $\rho$ grows, enhancing RLA's localization accuracy. Consequently, $\Delta{\rm TR}_2$ also increases. When $\rho=90^{\circ}$, as shown in Fig. \ref{fig:CRLB}(d), the intersection angle between $\calA_{\qo}$ and $\calA_{\qdelta}$  reaches its maximum, leading to VLA achieving its best localization performance. However, as $\rho$ continues to increase, as depicted in Figs. \ref{fig:CRLB}(e) and \ref{fig:CRLB}(f), the intersection angle between $\calA_{\qo}$ and $\calA_{\qdelta}$, now equals to $(180^{\circ}-\rho)$,  begins to decrease, causing VLA's localization performance to decline. In contrast, the intersection angle between $\calA$ and $\calD$ continues to increase, so $\Delta{\rm TR}_2$ keeps rising because RLA still has more additional information than VLA. As illustrated in Fig. \ref{fig:CRLB}(g), $\calA_{\qo}$ and $\calA_{\qdelta}$ coincide again at $\rho=180^{\circ}$, while the intersection angle between $\calA$ and $\calD$  is at its peak. Consequently, while VLA's localization capability vanishes, the range difference information inherent to RLA maximizes $\Delta{\rm TR}_2$.

\begin{figure}
\centering
\includegraphics[width=3.4in]{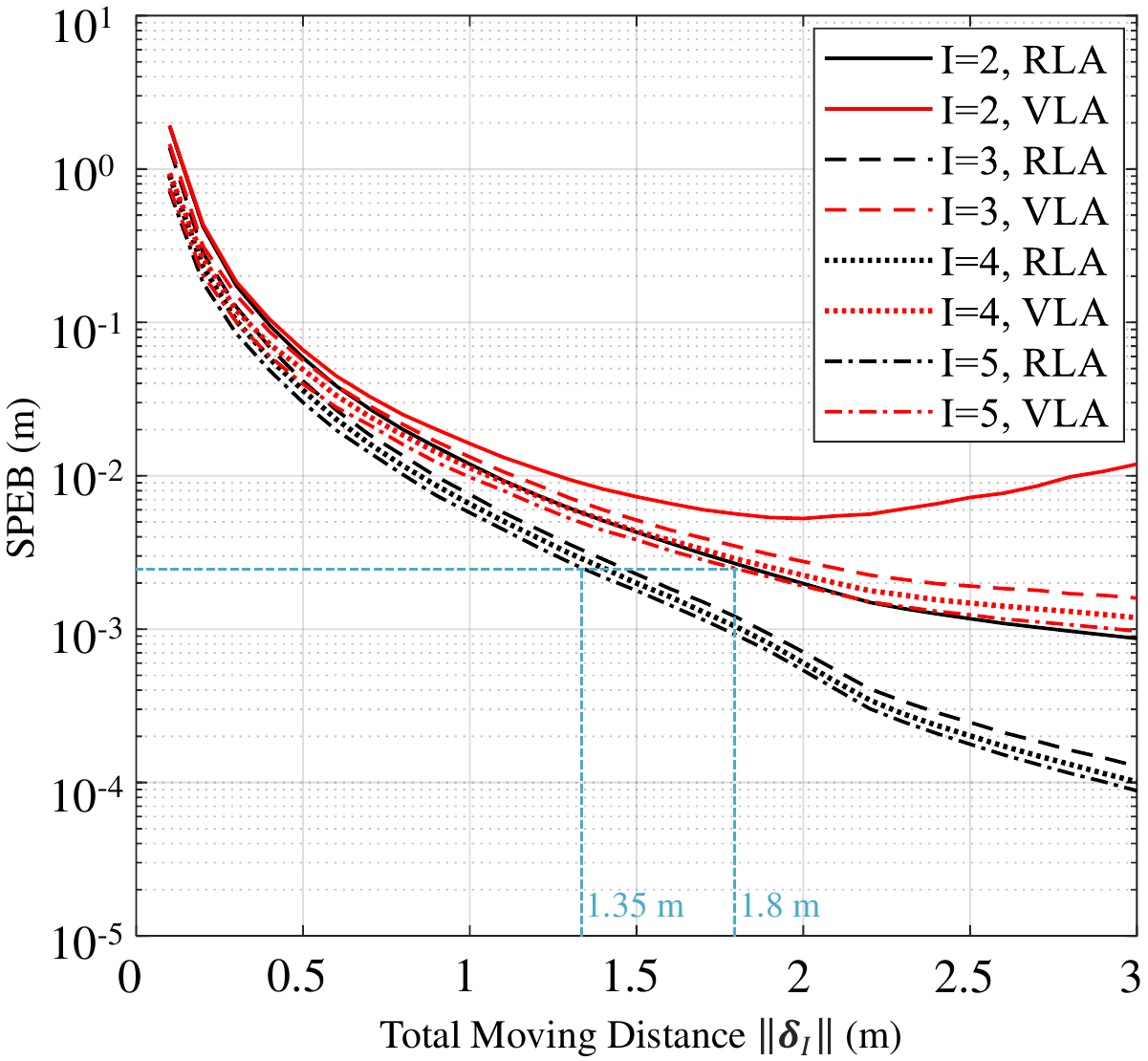}
\caption{The SPEB of RLA and VLA with different values of $\|\qdelta_I\|$.}
\label{fig:SPEB}
\end{figure}

To further explore these observations, we conduct a series of Monte Carlo simulations using IoT device 1 from Fig. \ref{fig:Layout}(a) as the transmitter, examining the more general cases with varying values of $I$. In each simulation, we model a person walking in a straight line in a random direction, naturally shaking the UE by hand to form a VLA. We ensure that the UE can capture LoS signals throughout all walking trajectories. Given that the movement of the UE caused by shaking is much smaller than that caused by walking, the UE's movement can essentially be considered linear. Additionally, both RLA and VLA use the same origin point $\qo$, sets of displacements ${\qdelta_1, \ldots, \qdelta_I}$, and orientations ${\qOmega_1, \ldots, \qOmega_I}$. For each distinct total movement distance $\|\qdelta_I\|$, we conduct 100,000 simulations. The SPEBs of RLA and VLA for various $\|\qdelta_I\|$ are plotted in Fig. \ref{fig:SPEB}. As $\|\qdelta_I\|$ increases, $\rho$ typically increases as well. Thus, when $I=2$, the SPEB of VLA first decreases and then increases, as VLA cannot accurately estimate the location when $\rho$ is close to $0^{\circ}$ and $180^{\circ}$. Conversely, the SPEB of RLA monotonically decreases as $\rho$ approaches $180^{\circ}$, benefiting from the additional range difference information. These observations align with our insights from Proposition \ref{th:D_FIM}.

For cases with $I = 3$, $4$, and $5$, the SPEBs of VLA show a monotonically decreasing trend with increasing $\|\qdelta_I\|$, contrary to the increase observed for $I = 2$. This pattern emerges because acquiring additional angle information enables VLA to determine the location through triangulation more precisely. Therefore, there is a notable improvement in SPEB for both VLA and RLA when $I > 2$, although this improvement tends to plateau beyond $I \geq 3$. As expected, VLA's localization performance lags behind RLA's. However, as shown in Fig. \ref{fig:SPEB} for ${I=5}$, when a person walks 1.8 m in three steps, with an average stride of 0.6 m, the VLA matches RLA's performance, which requires a 1.35-meter array aperture. Given the impracticality of integrating such an RLA into a UE, improving localization performance without requiring an RLA can be achieved through further UE movement and additional CFR captures.

\section{Localization Algorithm}

Before transitioning to the development of a practical localization algorithm, we note that traditional two-step localization methods cannot achieve outstanding performance because when errors are induced while extracting location-bearing parameters from the CFR, these errors will affect the localization results in the second step. In contrast, direct localization methods infer the sources' locations straightaway from the CFR, eliminating the unreliable intermediate step, thus surpassing the two-step process.

Considering that localizing the PS might be imprecise due to the interference caused by the VSs included in the total CFR, simultaneously estimating the locations of both the PS and VSs is necessary to enhance the PS's localization accuracy. Therefore, we utilize an optimization function:
   \begin{equation}
    \label{eq:arg_multipath}
    (\hat{\qalpha}, \hat{\qP}, \hat{\qtau}) = \argmin_{\qalpha, \qP, \qtau} \sum_{i=1}^{I} {\left\|  \qh_i - \sum_{l=1}^{L} { \alpha_{i,l} \qv(\qp_l, \tau_i;\,\qdelta_i, \qOmega_i) }  \right \|^2},
    \end{equation}
where $\qalpha = [\alpha_{1,1},\ldots,\alpha_{I,L}]^{\rm T}$, $\qP = [\qp_1,\ldots,\qp_L]$, and $\qtau = [\tau_1,\ldots,\tau_I]^{\rm T}$ are the parameters to be determined. Notably, VSs' location are estimated solely as auxiliary parameters to reduce multipath interference. However, seeking the optimal solution in the potentially vast domain of these parameters is a formidable task. To address this, we implement the framework of NOMP \cite{NOMP}.

NOMP iteratively estimates the parameters of each source. Each iteration of NOMP includes OMP, Newtonized refinement (NR), least squares (LS), and removal steps. The OMP step initially estimates the parameters of one source through a grid search method, which is then refined using Newton's method in the NR step. The LS and removal steps help ensure that the residual energy of the CFR is minimized after removing the estimated source's CFR.

The grid search method in the OMP step requires a specific search space to estimate the source's location. Conventionally, room size or other prior information is used to define this space. However, such information is not always available. Therefore, in this section, we propose ML-NOMP, a practical direct localization method that employs ML to initialize the search space for the PS, especially when prior information is lacking. Additionally, we use estimated distance to define the search space for the VSs. We introduce the NOMP estimation method in Section \ref{sec:NOMP_est}, provide details on determining the search space for PS and VSs in Sections \ref{sec:Search_LoS} and \ref{sec:Search_NLoS}, respectively, and summarize the entire process in Section \ref{sec:ML-NOMP_Summary}.

\subsection{Direct Localization via NOMP} \label{sec:NOMP_est}

To simplify the complex task of estimating the parameters of all sources simultaneously, NOMP first considers the estimation for one source and then expands to include multiple sources. Similar to \eqref{eq:VLA_channel}, we simplify \eqref{eq:arg_multipath} by removing the path index $l$, resulting in
    \begin{equation}
    \label{eq:arg_1}
    (\hat{\qalpha}, \hat{\qp}, \hat{\qtau}) = \argmin_{\qalpha, \qp, \qtau} \sum_{i=1}^{I} {\left\|  \qh_i - { \alpha_i \qv(\qp, \tau_i;\,\qdelta_i, \qOmega_i) }  \right \|^2},
    \end{equation}
where $\qalpha = [\alpha_1,\ldots,\alpha_I]^{\rm T}$. The solution to \eqref{eq:arg_1} can be obtained by maximizing the cost function:
    \begin{multline}
    \label{eq:cost_func}
    S(\qalpha, \qp, \qtau) = \sum_{i=1}^{I}  2\Re {\left\{ \alpha_i  \qv^{\rm H}(\qp, \tau_i;\,\qdelta_i, \qOmega_i) {\qh_i}  \right\}} \\
                                                                    - {\left|  \alpha_i \right|}^2 {\left\| \qv(\qp, \tau_i;\,\qdelta_i, \qOmega_i)\right\|}^2 .
    \end{multline}
For any given $\qp$ and $\tau_i$, the optimal $\alpha_i$ maximizing the $i$-th term in \eqref{eq:cost_func} is determined through LS estimation:
    \begin{equation}
    \label{eq:gain}
    \alpha_i = \frac{ \qv^{\rm{H}}(\qp, \tau_i;\,\qdelta_i, \qOmega_i) \qh_i }{\left\| \qv(\qp, \tau_i;\,\qdelta_i, \qOmega_i) \right\|^2}.
    \end{equation}
Substituting \eqref{eq:gain} into \eqref{eq:cost_func} leads to a new optimization problem:
    \begin{equation}
    \label{eq:arg2}
    (\hat{\qp}, \hat{\qtau})  = \argmax_{\qp, \qtau}  \sum_{i=1}^{I} {\frac{ \left|   \qv^{\rm{H}}(\qp, \tau_i;\,\qdelta_i, \qOmega_i) \qh_i   \right|^2}{\left\|  \qv(\qp, \tau_i;\,\qdelta_i, \qOmega_i) \right\|^2}}.
    \end{equation}

The problem in \eqref{eq:arg2} can be approached by grid searching through a finite feasible set $\calF_{\qp}$, created with location elements within an appropriate search space. The details of determining this space will be introduced in Sections \ref{sec:Search_LoS} and \ref{sec:Search_NLoS}. As IoT devices and UEs are typically not perfectly synchronized, estimating the time offset $\tau_i$ is also necessary. Thus, we form another set $\calF_\tau$ for each $\tau_i$, containing elements
    \begin{equation}
    \label{eq:subset_tau}
    \tau'  \in \calF_\tau \triangleq \left\{ k/(\eta_{\tau} W):  k = 0, 1, \ldots, \eta_{\tau} N-1 \right\},
    \end{equation}
where $W$ is the bandwidth, and $\eta_{\tau}$ is the sampling rate relative to the grid of $\tau$. The solution $(\hat{\qp}, \hat{\qtau})$ is found by substituting elements from $\calF_\qp$ and $\calF_\tau$ into \eqref{eq:arg2}, and $\hat\alpha_i$ is obtained by applying $(\hat{\qp}, \hat{\tau}_i)$ to \eqref{eq:gain} for each $i$.

An off-grid effect occurs if the elements of $\calF_{\qp}$ and $\calF_\tau$ are discrete. As searching over a continuum set to find the best solution is impractical, we refine the estimation from OMP using Newton's method. The refinement is repeated several times after the OMP step, and each refinement calculates
    \begin{equation}
    \label{eq:NR}
    \begin{bmatrix}   \hat{\qp}^{+}  \\  \hat{\qtau}^{+} \end{bmatrix} =
    \begin{bmatrix}   \hat{\qp}^{-}   \\  \hat{\qtau}^{-}  \end{bmatrix} -
    \ddot{S}^{-1}(\hat\qalpha^{-}, \hat\qp^{-}, \hat\qtau^{-}) \dot{S}(\hat\qalpha^{-}, \hat\qp^{-}, \hat\qtau^{-}),
    \end{equation}
where $\hat{\qp}^{-}$ and $ \hat{\qtau}^{-}$ are the results from the previous refinement, $\dot{S}(\hat\qalpha^{-}, \hat\qp^{-}, \hat\qtau^{-})$ and $\ddot{S}(\hat\qalpha^{-}, \hat\qp^{-}, \hat\qtau^{-})$ are the first and second partial derivatives, respectively, of \eqref{eq:cost_func} with respect to $ [\qp^{\rm T}, \qtau^{\rm T} ]^{\rm T}$. For the first refinement, $(\hat\qalpha^{-}, \hat\qp^{-}, \hat\qtau^{-})$ comes from the OMP estimation. Whenever we obtain $(\hat{\qp}^{+}, \hat{\tau}_i^{+})$ from \eqref{eq:NR}, we use $(\hat{\qp}^{+}, \hat{\tau}_i^{+})$ in \eqref{eq:gain} to refine $\hat{\alpha}_i^{+}$ for each $i$.

The estimation is expanded to multiple sources. At the $l$-th NOMP iteration, we obtain $[\hat\alpha_{1,l}, \dots, \hat\alpha_{I,l}]^{\rm T}$, $\hat\qp_{l}$, and $[\hat\tau_{1}, \dots, \hat\tau_{I}]^{\rm T}$ by OMP and NR steps. Since an estimate of one source might be influenced by the CFR caused by others, the estimates of the first to the $l$-th sources are cyclically refined $K^{\rm NR}$ times using NR. After this cyclic NR step, we ensure maximizing the power of the estimated sources through the LS step:
    \begin{equation}
    \label{eq:LS}
    \hat\qalpha_i = \left( \qV_i^{\rm H} \qV_i \right)^{-1} \qV_i^{\rm H} \qh_i,
    \end{equation}
where $\hat\qalpha_i = \left[ \hat\alpha_{i,1},\ldots,\hat\alpha_{i,l} \right]^{\rm T}$, and
    \begin{equation}
    \label{eq:V}
    \qV_i = [\qv(\hat{\qp}_1, \hat\tau_i;\,\qdelta_i, \qOmega_i), \ldots, \qv(\hat{\qp}_{l}, \hat\tau_i;\,\qdelta_i, \qOmega_i)]
    \end{equation}
is the combination of the CFR vectors estimated from the first to the $l$-th NOMP iteration. In the final step, the residual CFR is calculated by removing the CFR of the estimated sources:
    \begin{equation}
    \label{eq:remove}
    \qh_i^{\rm res} = \qh_i - \sum_{l'=1}^{l} { \hat\alpha_{i,l'} \qv(\hat{\qp}_{l'}, \hat\tau_i;\,\qdelta_i, \qOmega_i) },
    \end{equation}
and $\qh_i^{\rm res}$ is then used in \eqref{eq:arg2} for the next NOMP iteration.

\begin{figure*}
\centering
\includegraphics[width=7.0in]{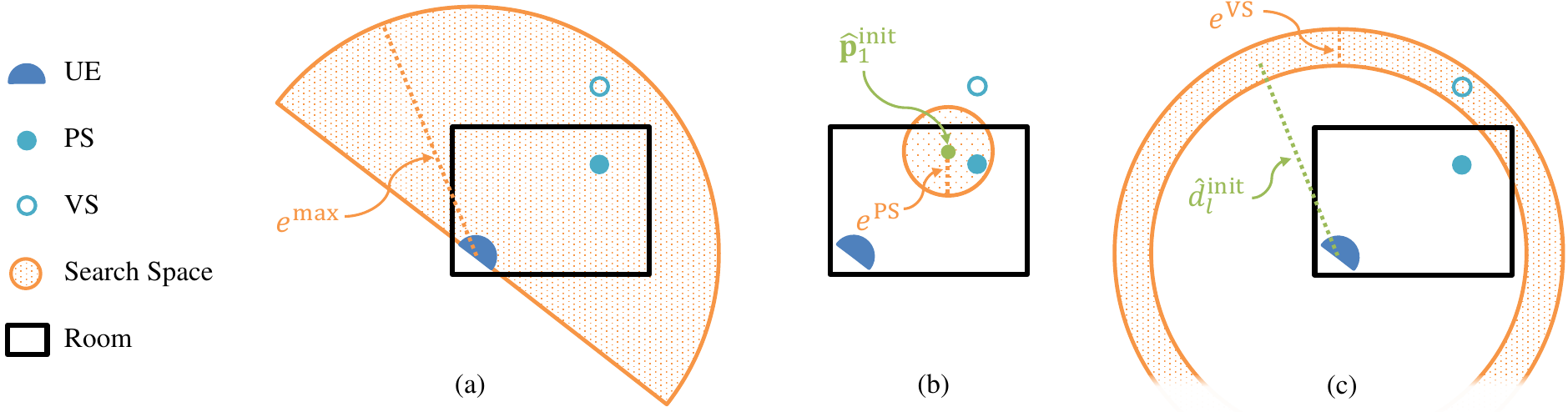}
\caption{Search space determined for different sources by various methods: (a) applying the traditional method for PS, (b) applying the ML method for PS, (c) utilizing estimated distance for VS.}
\label{fig:Searching Space}
\end{figure*}

\subsection{Search Space Initialization for PS} \label{sec:Search_LoS}

\begin{figure*}
\centering
\includegraphics[width=7.0in]{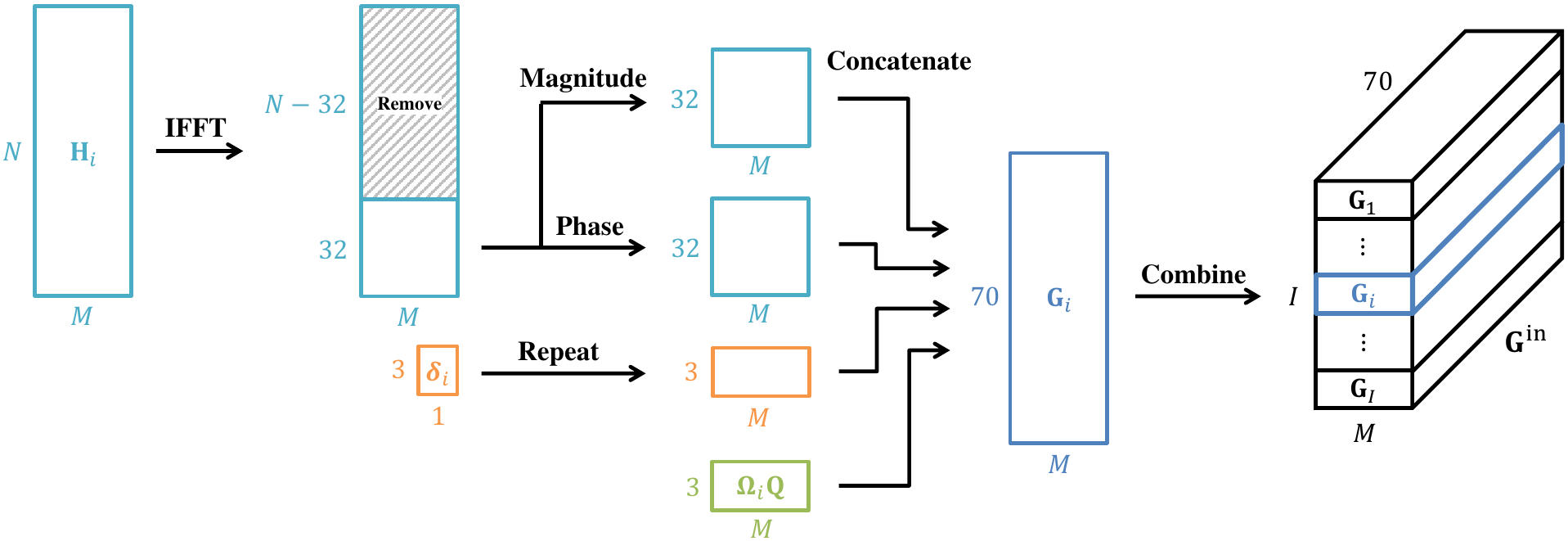}
\caption{Input data for a single trajectory.}
\label{fig:ML_input}
\end{figure*}

As indicated in Section \ref{sec:NOMP_est}, it is essential to have appropriate search spaces to determine the location set $\calF_{\qp}$. These search spaces vary for each source. This subsection primarily focuses on the search space for the PS. Traditionally, we assume prior knowledge about the room, such as its dimensions (e.g., ${A \times B \times C}$), is obtainable. Given that the UE and IoT device can be placed anywhere in the room, the maximum distance between them, $e^{\rm{max}} = \sqrt{A^2 + B^2 + C^2}$, limits the search space. As depicted in Fig. \ref{fig:Searching Space}(a), we create a finite feasible set $\calF_{\qp}= \{(d', \phi', \theta')\}$ with elements defined as
    \begin{align}
    \label{eq:traditional_subset}
    d'       &\in \calF_{d}        = \left\{e^{\rm{max}}/\eta_{d} \cdot k:  k = 0, 1, \ldots, \eta_{d} \right\}, \notag\\
    \phi'    &\in \calF_{\phi}    = \left\{2\pi/\eta_{\phi} \cdot k-\pi:          k = 0, 1, \ldots, \eta_{\phi} \right\},        \\
    \theta'  &\in \calF_{\theta} = \left\{\pi/\eta_{\theta} \cdot k:              k = 0, 1, \ldots, \eta_{\theta} \right\}\notag,
    \end{align}
where $\eta_{d}$, $\eta_{\phi}$, and $\eta_{\theta}$ are the sampling rate relative to the grid of distance, elevation, and azimuth, respectively. Due to the UE's patch antennas having boresight in front, azimuth coordinates $(\pi, 2\pi)$ are excluded. The set $\calF_{\qp}= \{ (d', \phi', \theta') \}$ corresponds to the position $\{ (x', y', z') \}$ following \eqref{eq:spherical_model}.

The traditional method faces a challenge: obtaining the room layout is not always feasible in practice. Moreover, computational time increases with $e^{\rm{max}}$. To address this, we apply ML techniques to derive a search space for the PS. By inputting the CFR and sensor information into a trained ML model, we obtain an approximate PS location and generate a smaller set $\calF_{\qp}$ around this estimation.

For ML model training, we collect a database as the UE moves along different trajectories. On each trajectory, we gather the CFR $\qh_i$, movement $\qdelta_i$, and the rotated antennas' positions relative to the UE's center $\qOmega_i\qQ$ for every $i$, where $\qQ=[\qq_1,\ldots,\qq_M]$. As illustrated in Fig. \ref{fig:ML_input}, we transform the CFR vector $\qh_i$ into a matrix form $\qH_i$, with the $m$-th column containing the $m$-th antenna's CFR, and convert each column into time-domain channel impulse responses (CIRs) using the inverse fast Fourier transform (IFFT). Focusing on LoS estimation, we retain the first 32 points\footnote{The number of points retained is based on our experiments and can be adjusted for different systems.} of CIR to preserve the PS's CIR and minimize multipath effects. The magnitude and phase of CIRs are calculated, yielding two $32 \times M$ matrices. To incorporate sensor information, we repeat the vector $\qdelta_i$ $M$ times, forming a $3 \times M$ matrix. Each moment's information is concatenated into $\qG_i \in \mathbb{R}^{70 \times M}$, and $\qG_i$ for $i = 1, \ldots, I$ are combined into a 3D array $\qG^{\rm in} \in \mathbb{R}^{70 \times I \times M}$.

\begin{figure}
\centering
\includegraphics[width=3.5in]{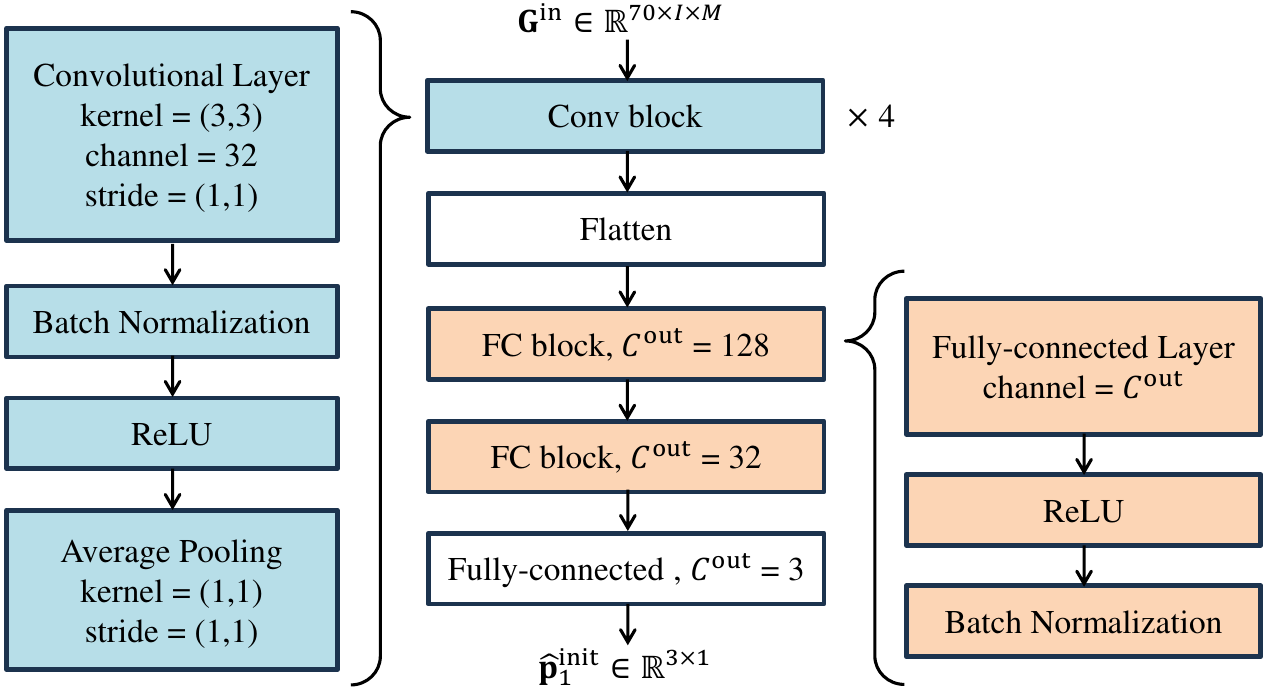}
\caption{The CNN for $I=7$ and $M=3$.}
\label{fig:ML_model}
\end{figure}

Employing a larger ML model with extensive training data could potentially enhance the model's robustness. However, this approach may also result in increased complexity and higher power consumption, which are critical considerations for UEs such as smartphones. Additionally, the performance of the ML model is not always guaranteed. Therefore, we opt for a lightweight convolutional neural network (CNN) that is easier to train and deploy on smartphones, as shown in Fig. \ref{fig:ML_model}. The proposed neural architecture consists of four convolutional layers, each with 32 filters of size $3 \times 3$, followed by two fully connected layers with 128 and 32 neurons, respectively, designed to produce a $3 \times 1$ vector output representing the PS's x-, y-, and z-coordinates:
    \begin{equation}
    \label{eq:ML_est}
    \hat{\qp}_1^{\rm init} = K(\qG^{\rm in};\Theta),
    \end{equation}
where $K(\cdot)$ is the model function, and $\Theta$ represents the learned weights. The loss function, defined as the mean square error (MSE) of the location, is
    \begin{equation}
    \label{eq:loss_function}
    {\rm MSE} = \frac{1}{T} \sum_{t=1}^{T} {\left\| \qp_{1}[t] - \hat{\qp}^{\rm init}_{1}[t] \right\|},
    \end{equation}
where $\qp_{1}[t]$ and $\hat{\qp}^{\rm init}_{1}[t]$ are the ground truth and initial estimated PS location for the $t$-th training sample, respectively, and $T$  is the total number of training samples. This ML model is trained using the Adam optimizer with a batch size of 1,000.

Notably, the input data, $\qG^{\rm in}$, does not include any information about the room size. We directly estimate the PS's location with the trained ML model. However, when training and testing data come from different rooms or environments, the localization accuracy might degrade. To mitigate this issue, we use the estimated location, $\hat{\qp}^{\rm init}_1$, as a starting point and create a search space around it, as depicted in Fig. \ref{fig:Searching Space}(b). Specifically, we generate a set $\calF_{\qp_1}$ with elements:
    \begin{align}
    \label{eq:ML_subset}
    d'       &\in \calF_{d_1}        = \left\{e^{\rm{PS}}/\eta_{d} \cdot k: k = 0, 1, \ldots, \eta_{d} \right\}, \notag\\
    \phi'    &\in \calF_{\phi_1}    = \left\{2\pi/\eta_{\phi} \cdot k-\pi:       k = 0, 1, \ldots, \eta_{\phi} \right\}, \\
    \theta'  &\in \calF_{\theta_1} = \left\{2\pi/\eta_{\theta} \cdot k:         k = 0, 1, \ldots, \eta_{\theta}-1 \right\}\notag,
    \end{align}
where $e^{\rm PS}$ is a smaller range than $e^{\rm max}$. By centering these elements around $\hat{\qp}^{\rm init}_1$, we form the feasible set $\calF_{\qp_1}$, which can significantly reduce computation time compared to the traditional search space defined in \eqref{eq:traditional_subset}.

\subsection{Search Space Initialization for VSs} \label{sec:Search_NLoS}

In the previous subsection, we discussed initializing the search space for PSs using traditional and advanced ML methods. However, these methods are not directly applicable to restrict the search space for VSs due to unpredictable obstacles within the room causing diverse reflections and refractions. Additionally, VSs' characteristics are hard to learn from raw CIR data, as a VS may cause different propagation delays at different moments due to the changing distance between the UE and the VS as the UE moves.

To tackle the challenge of initializing the search space for VSs, we focus primarily on $\qh_1$, the CFR captured at the initial moment. We define the propagation delay of $\qp_l$ coupled with the unknown time offset $\tau_1$ in \eqref{eq:subcarrier_response_general} as \emph{the relative delay}
    \begin{equation}
    \label{eq:ToA_with _offset}
    \tilde{\tau}_{1,l} = \frac{\| \qp_l \|}{c} - \tau_1.
    \end{equation}
After completing the ${(l-1)}$-th iteration of the NOMP, we compute the residual CFR, denoted as $\qh_1^{\rm res}$. At the onset of the $l$-th NOMP iteration, we apply the IFFT on $\qh_1^{\rm res}$ to derive the CIR of the residual VSs. Given that the first $(l-1)$ CIRs have been removed, the CIR exhibiting the largest peak is identified as stemming from $\qp_l$, allowing us to observe the relative delay $\tilde{\tau}_{1,l}$ from this CIR. By substituting $\hat{\tau}_1$, the estimated time offset from the preceding NOMP iteration, with $\tau_1$ in \eqref{eq:ToA_with _offset}, we ascertain the actual propagation delay, $\|\qp_l\| / c$, and subsequently, the initial distance $\hat{d}_l^{\rm init}$ for the $l$-th VS can be calculated by multiplying this estimated propagation delay by the speed of light, $c$. As depicted in Fig. \ref{fig:Searching Space}(c), we establish the feasible set $\calF_{\qp_l}$ based on this initially computed distance, with elements defined as
    \begin{align}
    \label{eq:ToA_subset}
    d'       &\in \calF_{d_l}       = \left\{e^{\rm VS}/\eta_{d} \cdot k + \hat d_l^{\rm init}:   k = -\eta_{d}, -\eta_{d}+1, \ldots,  \eta_{d} \right\}, \notag\\
    \phi'    &\in \calF_{\phi_l}   = \left\{2\pi/\eta_{\phi} \cdot k-\pi:                                     k = 0, 1, \ldots, \eta_{\phi} \right\}, \\
    \theta' &\in \calF_{\theta_l} = \left\{2\pi/\eta_{\theta} \cdot k:                                        k = 0, 1, \ldots, \eta_{\theta}-1 \right\}\notag,
    \end{align}
where $e^{\rm VS}$ is a minor range around the estimated distance. Then, we employ the same NOMP framework introduced in Section \ref{sec:NOMP_est} to pinpoint the $l$-th VSs' location, $\hat{\qp}_l$.

\subsection{Summary of ML-NOMP Algorithm} \label{sec:ML-NOMP_Summary}

In conclusion, estimating the positions of the PS and VSs can increase localization accuracy. Therefore, we apply the NOMP, an iterative method, to reduce the complexity of estimating all the sources simultaneously. Moreover, for the first NOMP iteration, we propose ML methods to determine the PS's search space. The time offset estimated from the first NOMP iteration can then derive the VSs' distance and search space starting from the second NOMP iteration. After defining the sources' search space, we estimate and refine the sources' location through OMP and NR steps, and minimize the power of the residual CFR by LS and removal steps before delivering the residual CFR to the next NOMP iteration. The entire process is summarized in Algorithm \ref{alg:ML-NOMP}.

\begin{algorithm}
\label{alg:ML-NOMP} \small
\caption{ML-NOMP}
\KwIn{$\qh_i, \qdelta_i, \qOmega_i$ for $i=1, \ldots, I$}
\KwOut{$\hat\qP = [\hat{\qp}_0, \hat{\qp}_1, \ldots, \hat{\qp}_ L]$.}
\For{$l= 1,\ldots,L$}{
    {\bf Initialization}:
            Create feasible set $\calF_{\tau}$ by \eqref{eq:subset_tau} and \\
            \eIf{$l = 1$}
                    {   create feasible set $\calF_{\qp_1}$ by \eqref{eq:ML_est} and \eqref{eq:ML_subset}\;   }
                    {   create feasible set $\calF_{\qp_l}$ by \eqref{eq:ToA_subset}\;  }
    {\bf OMP}:
            Solve $\hat{\qp}_l$ and $\hat\qtau=[\hat\tau_{1},\ldots,\hat\tau_{I}]^{\rm T}$ by substituting $\calF_{\qp_l}$ and $\calF_{\tau}$ into \eqref{eq:arg2}, and obtain $\hat{\qalpha}=[\hat\alpha_{1,l},\ldots,\hat\alpha_{I,l}]^{\rm T}$ by substituting $(\hat{\qp}_l, \hat\tau_i)$ into \eqref{eq:gain} for all $i$\;
    {\bf NR}:
            Set the temporary results $(\hat{\qalpha}, \hat{\qP}, \hat\qtau)$ as $(\hat{\qalpha}^{-}, \hat{\qP}^{-}, \hat\qtau^{-})$ and \\
            \For{$k'= 1,\ldots, K^{\rm NR}$}{
                    \For{$l'= 1,\ldots, l$}{
                            Refine $(\hat{\qp}_{l'}^{-},\hat\qtau^{-})$ by \eqref{eq:NR} to obtain $(\hat{\qp}_{l'}^{+},\hat\qtau^{+})$, and refine $\qalpha_{i,l'}^{+}$ by taking $(\hat{\qp}_{l'}^{+},\hat\qtau_{i}^{+})$ into \eqref{eq:gain}\;
                    }
            }
    {\bf LS}:
            Calculate $\hat\qalpha_i$ by \eqref{eq:LS} for all $i$\;
    {\bf Remove}:
            Calculate $\qh_i^{\rm res}$ by \eqref{eq:remove} for all $i$\;
}
\end{algorithm}

It is important to note that initial estimations of $\hat{\qp}_1^{\rm init}$ may be affected by the CFR of the VSs, and the small range $e^{\rm PS}$ might not cover the ground truth position, leading to potential inaccuracies in creating the search space for estimating $\hat{\qp}_1$ and $\hat{\tau}_1$. Although an inaccurate $\hat{\tau}_1$ causes an error on $\hat d_2^{\rm init}$ and the small range $e^{\rm VS}$ cannot determine a suitable search space for estimating precise $\qp_2$, the following cyclic NR step can refine $\hat{\qp}_1$, $\hat{\qp}_2$, and $\hat{\tau}_1$. Consequently, the refined $\hat{\tau}_1$ enables a better search space for the third iteration, allowing for a more accurate estimation of $\qp_3$, which in turn helps refine $\hat{\qp}_1$, $\hat{\qp}_2$, $\hat{\qp}_3$, and $\hat{\tau}_1$ during the cyclic NR step.

Although the search spaces with small range $e^{\rm PS}$ and $e^{\rm VS}$ can reduce computation complexity, they can also have a bias based on an inaccurate initial ML estimation. However, the NR step facilitates obtaining a more effective search space for the next iteration, and an improved search space, in turn, enhances the accuracy of the estimates in the NR step. Finally, the error caused by ML can be revised.

\section{Simulation Results}

\subsection{Setup and SOTA}

We conduct 3D ray-tracing simulations using Wireless InSite$\circledR$ on a personal computer equipped with Intel$\circledR$ Xeon$\circledR$ E-2124G CPU to evaluate the localization performance of ML-NOMP. The computer is also equipped with an NVIDIA$\circledR$ GeForce$\circledR$ RTX 2080 Ti GPU, which handles all calculations related to ML. The IoT device transmits an OFDM-based waveform at a central frequency of 6.5\,GHz with a channel bandwidth of 500\,MHz, incorporating $N = 1,644$ active subcarriers in one OFDM symbol.
The UE is equipped with a 2D x-z array:
    \begin{equation*}
    [\qq_1, \qq_2, \qq_3] = \frac{ \lambda}{2}
    \begin{bmatrix}
    -1/2 & 1/2 &   1/2      \\
       0  &   0   &    0        \\
    1/2  &  1/2 &  -1/2
    \end{bmatrix},
    \end{equation*}
where the patch antennas on this array receive signals only from azimuth angles ${\theta < \pi}$.


\begin{figure*}
\centering
\includegraphics[width=7.2in]{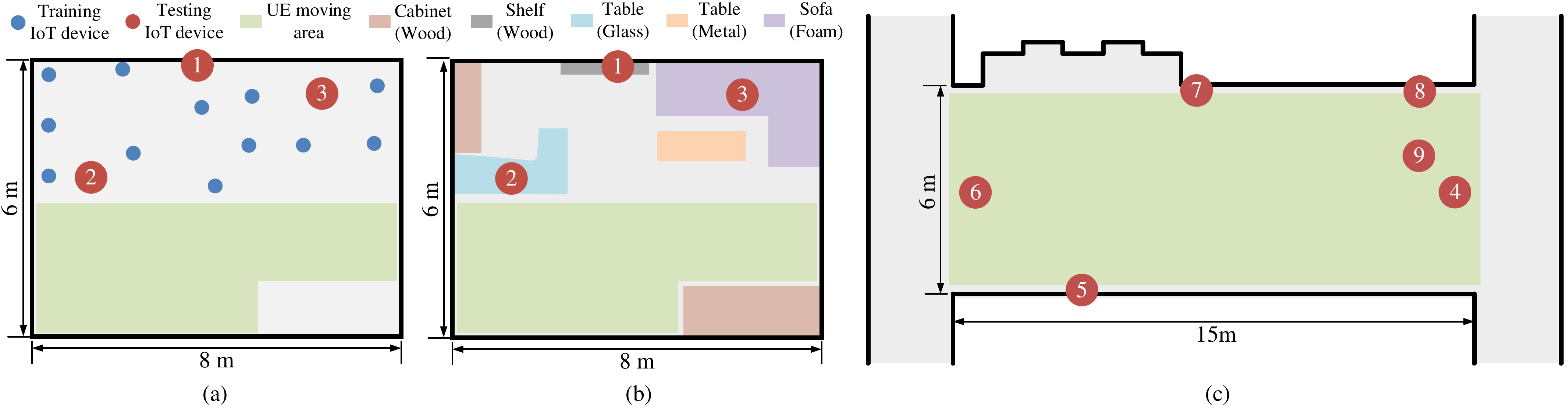}
\caption{Plan view of the simulation environment: (a) an $8\,{\rm{m}}\times6\,{\rm{m}}\times3\,{\rm{m}}$ empty room, (b) the same room as (a) but with added furniture, (c) a $15\,{\rm{m}}\times6\,{\rm{m}}\times3\,{\rm{m}}$  corridor at National Sun Yat-sen University, Taiwan.}
\label{fig:Layout}
\end{figure*}

Seven different SOTA methods are compared with ML-NOMP, including two-step LS, two-step SLAM \cite{2step_EasyAPPos}, OMP \cite{SOTA_RIS_NLoS}, NOMP \cite{SOTA_ELAA2}, Nelder-Mead (NM) \cite{SOTA_NM}, ML \cite{ML_RLA3}, and ML-NM. Two-step LS and two-step SLAM represent the basic and SOTA two-step localization methods, respectively. Both two-step methods use NOMP to estimate the AoA by first performing a coarse grid search and then refining the estimate using Newton's method. In the second step, LS and SLAM integrate the AoA estimation with IMU information. OMP and NOMP, introduced in Section \ref{sec:NOMP_est}, are also evaluated. OMP estimates only the PS's location by substituting \eqref{eq:subset_tau} and \eqref{eq:traditional_subset}, which consider the widest range based on the room size, into the optimal function \eqref{eq:arg2}. NOMP iteratively estimates both PS's and VSs' locations using OMP and cyclic NR steps, with the performance presented for different iteration counts, denoted as $L$. ML estimates only the PS's location using the trained CNN shown in Fig. \ref{fig:ML_model}; unlike OMP, ML does not rely on prior information about the room size. NM and ML-NM are similar to NOMP and ML-NOMP, respectively, but they replace NR with the Nelder-Mead refinement (NMR) after the coarse estimation. Other relevant parameter settings are $\eta_{\tau}=8$, $\eta_d=4$, $\eta_{\phi} = \eta_{\theta} = 120$, $e^{\rm PS}=1.2\,{\rm m}$, and $K^{\rm NR}=3$, which means the NR steps are repeated 3 times.

The simulations model an indoor scenario where a person takes three steps forward in a straight trajectory with an average pace of 0.6\,m, resulting in a total movement of 1.8\,m. It is assumed that a LoS path always exists between the UE and the PS. The CFRs and sensor information captured by the UE at seven moments are combined, as illustrated in Fig. \ref{fig:ML_input}. We randomly place 12 IoT devices at different heights in the room layout shown in Fig. \ref{fig:Layout}(a), generating 90,000 training data points and 10,000 validation data points for each IoT device to train the CNN model depicted in Fig. \ref{fig:ML_model}.

\subsection{Testing in an Empty Room}
\label{subsec:empty}

\begin{table*}[!t]
    \vspace{0.5cm}
    \begin{minipage}{.5\textwidth}
        \centering
        \caption{Average localization error (m) of different localization methods tested in the environment shown in Fig. \ref{fig:Layout}(a).}
        \label{tab:results_A}
        \begin{tabular}{ l c c c c c } \toprule
    & {\multirow{2}{*}{\makecell*[c]{Avg. run\\time (ms)}}}      & \multicolumn{4}{c}{Avg. localization error (m)}       \\ \cmidrule(lr){3-6}
    &                                                               & IoT 1      & IoT 2         & IoT 3        & Total            \\ \midrule
                                                 Two-step LS & {\bf 2.2} & 0.92         & 0.53           & 1.23          & 0.89             \\ \midrule
    Two-step SLAM\cite{2step_EasyAPPos} & 81.4       & 0.80          & 0.55           & 1.15          & 0.83             \\ \midrule
                    OMP \cite{SOTA_RIS_NLoS} & 9.8         & 0.98          & 0.52           & 0.84          & 0.78             \\ \midrule
          NOMP \cite{SOTA_ELAA2}, $L=1$ & 12.5       & 0.86          & 0.45           & 0.79          & 0.70             \\ \midrule
          NOMP \cite{SOTA_ELAA2}, $L=2$ & 19.0       & 0.77          & 0.43           & 0.77          & 0.66             \\ \midrule
          NOMP \cite{SOTA_ELAA2}, $L=3$ & 27.3       & 0.74          & 0.40           & 0.75          & 0.63             \\ \midrule
                    NM \cite{SOTA_NM}, $L=1$ & 39.8        & 0.60          & 0.36           & 0.61          & 0.52              \\ \midrule
                                  ML \cite{ML_RLA3} & 2.9          & 0.43          & 0.31           & {\bf 0.36} & 0.37             \\ \midrule
                                     ML-NOMP, $L=1$ & 6.2          & 0.53          & 0.31           & 0.46          & 0.43             \\ \midrule
                                     ML-NOMP, $L=2$ & 12.7        & 0.42          & 0.30           & 0.44          & 0.39             \\ \midrule
                                     ML-NOMP, $L=3$ & 21.0        & {\bf 0.40} & {\bf 0.27}  & 0.42          & {\bf 0.36}     \\ \midrule
                                          ML-NM, $L=1$ & 26.8        & 0.47          & 0.30           & 0.44          & 0.40              \\ \bottomrule
    \end{tabular}
    \end{minipage}
    \begin{minipage}{.5\textwidth}
        \centering
        \caption{Average localization error (m) of different localization methods tested in the environment shown in Fig. \ref{fig:Layout}(b).}
        \label{tab:results_B}
        \begin{tabular}{ l c c c c c } \toprule
    & {\multirow{2}{*}{\makecell*[c]{Avg. run\\time (ms)}}}       & \multicolumn{4}{c}{Avg. localization error (m)}     \\ \cmidrule(lr){3-6}
    &                                                                                   & IoT 1       & IoT 2        & IoT 3           & Total         \\ \midrule
                                                  Two-step LS & {\bf 2.2} & 0.77        & 0.60          & 1.85            & 1.07          \\ \midrule
    Two-step SLAM\cite{2step_EasyAPPos} & 81.4        & 0.74        & 0.54          & 1.41            & 0.90          \\ \midrule
                    OMP \cite{SOTA_RIS_NLoS} &  9.8         & 0.90        & 0.52          & 1.21            & 0.88           \\ \midrule
          NOMP \cite{SOTA_ELAA2}, $L=1$ & 12.5        & 0.69        & 0.43          & 1.15            & 0.76           \\ \midrule
          NOMP \cite{SOTA_ELAA2}, $L=2$ & 19.0        & 0.60        & 0.41          & 1.14            & 0.71           \\ \midrule
          NOMP \cite{SOTA_ELAA2}, $L=3$ & 27.3        & 0.60         & 0.38          & 1.14           & 0.70           \\ \midrule
                    NM \cite{SOTA_NM}, $L=1$ & 39.8         & 0.48        & 0.36          & 1.03            & 0.62           \\ \midrule
                                   ML \cite{ML_RLA3} & 2.9          & 0.51         & 0.30          & 0.61           & 0.47            \\ \midrule
                                     ML-NOMP, $L=1$ & 6.2           & 0.50         & 0.32          & 0.62           & 0.47            \\ \midrule
                                     ML-NOMP, $L=2$ & 12.7         & 0.41         & 0.30          & 0.61           & 0.43             \\ \midrule
                                     ML-NOMP, $L=3$ & 21.0         & 0.41         & {\bf 0.27} & 0.55           & 0.41             \\ \midrule
                                          ML-NM, $L=1$ & 26.8         &{\bf 0.39} & 0.32          & {\bf 0.47}  & {\bf 0.39}   \\ \bottomrule
    \end{tabular}
    \end{minipage}
\end{table*}

For testing, we use the 3 IoT devices depicted in Fig. \ref{fig:Layout}(a), generating 1,000 test datasets for each IoT device. The average running time and localization errors are summarized in Table \ref{tab:results_A}, with the shortest running times and lowest localization errors highlighted in bold. Notably, two-step LS has the shortest running time since the grid search for the AoA is much simpler than a grid search for the location. However, it exhibits the worst localization accuracy because errors in AoA estimation during the first step adversely affect the second localization step. Two-step SLAM improves upon the localization performance of two-step LS but requires significantly more time due to using a particle filter. Moreover, both two-step methods underperform the direct localization methods. Consequently, two-step localization methods may not offer advantages in terms of either computational time or localization accuracy when compared to direct localization approaches.

OMP leverages prior knowledge of room size to define a search space, as shown in \eqref{eq:traditional_subset}, with $e^{\rm max} = 10.44\,{\rm m}$. Exhaustively searching for the PS location within this extensive space takes 9.8\,ms. NOMP further refines OMP’s estimate using NR, improving localization performance as the number of iterations increases. NM refines the OMP estimate with one iteration of NMR, achieving better results than NOMP, but it requires significantly more time to converge.

Compared to OMP, ML requires no prior information and estimates the PS's location in just 2.9\,ms using a lightweight model. Additionally, ML outperforms OMP. This improvement is attributed to the ML model being trained and tested in the same environment, albeit at different IoT device positions, allowing it to learn additional environmental characteristics from the CFR. We also apply NR and NMR to ML, respectively. ML-NOMP refines the ML estimate through several iterations. Initially, its performance may be hindered by potential multipath interference affecting the NR steps, requiring more iterations to reduce the error and eventually surpass the performance of ML. On the contrary, ML-NM does not effectively refine the ML estimate, and it also requires significantly longer computation times.

\subsection{Testing in a Room with Furniture}

In Table \ref{tab:results_A}, the performance of ML appears comparable to that of ML-NOMP and ML-NM, suggesting that the refinement steps after ML might be redundant. However, ML's performance can degrade when the environment changes. To test the generalizability of ML, we modify the room layout by adding furniture, as illustrated in Fig. \ref{fig:Layout}(b). The furniture affects the CFRs due to complex penetrations and reflections. For instance, the shelf under IoT device 1 causes constructive interference in LoS and reflection paths, the metal table blocks the LoS signal between IoT device 3 and the UE, and the glass table under IoT device 2 does not significantly affect penetration and reflection.

This alteration does not change the range of the search space or the running time, but it does lead to different localization performances, as shown in Table \ref{tab:results_B}. Compared to Table \ref{tab:results_A}, the constructive interference improves the performance of two-step LS, two-step SLAM, OMP, NOMP, and NM on IoT device 1, while the attenuated LoS signal degrades the performance of these methods on IoT device 3. ML's performance consistently degrades with the changing CFRs, regardless of the specific impact of furniture on individual IoT devices. Nonetheless, both ML-NOMP and ML-NM can refine the estimates from ML, compensating for ML's lack of generalizability and enhancing localization accuracy.

\subsection{Testing in a Larger Corridor}

\begin{table*}[!t]
\caption{Average localization error (m) of different localization methods tested in the environment shown in Fig. \ref{fig:Layout}(c).  \label{tab:results_C}}
\centering
\begin{tabular}{ l c c c c c c c c} \toprule
    & {\multirow{2}{*}{\makecell*[c]{Avg. run\\time (ms)}}}       &\multicolumn{7}{c}{Avg. localization error (m) }                                                                    \\ \cmidrule(lr){3-9}
    &                                                                                            & IoT 4         & IoT 5        & IoT 6        & IoT 7            & IoT 8       & IoT 9        & Total       \\ \midrule
                                              Two-step LS   & {\bf 2.2} & 2.87          & 1.98          & 3.74           & 0.72         & 1.88          & 2.49           & 2.28  \\ \midrule
    Two-step SLAM\cite{2step_EasyAPPos}& 81.4        & 2.37          & 1.69          & 3.15           & 0.98          & 1.55          & 2.22           & 1.99  \\ \midrule
                   OMP \cite{SOTA_RIS_NLoS} & 15.9        & 1.49          & 1.30          & 2.01           & 0.95          & 1.48          & 1.53           & 1.46  \\ \midrule
         NOMP \cite{SOTA_ELAA2}, $L=1$ & 18.7        & 1.34          & 1.07          & 1.84           & 0.74          & 1.21          & 1.39           & 1.26  \\ \midrule
         NOMP \cite{SOTA_ELAA2}, $L=2$ & 25.2        & {\bf 1.30} & 1.05          & 1.80           & 0.69          & 1.18          & 1.36           & 1.23  \\ \midrule
         NOMP \cite{SOTA_ELAA2}, $L=3$ & 33.6        & {\bf 1.30} & {\bf 1.04} & 1.78           & 0.65          & 1.16          & 1.35           & 1.21  \\ \midrule
                   NM \cite{SOTA_NM}, $L=1$  & 57.0        & 1.31          & 1.05          & 1.78           & {\bf 0.60} & 1.17          & 1.25           & 1.19  \\ \midrule
                                  ML \cite{ML_RLA3} & 2.9          & 3.20          & 1.77          & 3.23           & 1.26           & 3.12          & 2.67          & 2.54  \\ \midrule
                                     ML-NOMP, $L=1$ & 6.2          & 1.79          & 1.23          & 2.10           & 0.86           & 1.52          & 1.49          & 1.50  \\ \midrule
                                     ML-NOMP, $L=2$ & 12.7        & 1.43          & 1.20          & 1.82           & 0.81           & 0.99          & 1.25          & 1.25  \\ \midrule
                                     ML-NOMP, $L=3$ & 21.0        & 1.32          & 1.21          & {\bf 1.71}  & 0.75           & {\bf 0.93} & {\bf 1.16} & {\bf 1.18} \\ \midrule
                                           ML-NM, $L=1$ & 44.0       & 2.83          & 1.29          & 2.90           & 0.66           & 2.79          & 2.27          & 2.12   \\ \bottomrule
\end{tabular}
\end{table*}

According to the results in Tables \ref{tab:results_A} and \ref{tab:results_B}, NOMP and NM improve localization accuracy by less than 0.3\,m compared to OMP, while ML-NOMP and ML-NM improve accuracy by less than 0.1\,m compared to ML. This is primarily because the sources in the small room depicted in Fig. \ref{fig:Layout}(a) and \ref{fig:Layout}(b) are close to the VLA, allowing simpler methods like OMP and ML to achieve accurate localization. Furthermore, the localization accuracy and computational time of NR with three iterations and NM with one iteration are nearly equal, making it difficult to determine which refinement method is superior. To better evaluate the advantage of applying NR or NMR to ML, we conduct simulations in a larger environment, as shown in Fig. \ref{fig:Layout}(c).

The results in Table \ref{tab:results_C} indicate that the different environment and increased distance between the sources and VLA significantly reduce ML's performance, leading to errors averaging 2.54\,m. However, ML still narrows the search space from 16.4\,m to 2.54\,m. ML-NM does not significantly improve localization accuracy because finding the optimal solution with a poor initial estimate from ML is challenging, and it requires more time for convergence. In contrast, ML-NOMP can significantly reduce errors through NR steps. Additionally, while the running time of OMP increases with room size, ML maintains a consistent running time since $e^{\rm PS}$ in \eqref{eq:ML_subset} remains unchanged. Notably, ML-NOMP, with just one iteration calculated in only 6.2\,ms, can improve accuracy to 1.5\,m and surpass the performance of OMP, which requires 15.9\,ms for an exhaustive grid search. Moreover, with three iterations completed in just 21\,ms, ML-NOMP can outperform NM, which requires 57\,ms for convergence.

\subsection{Testing in an Office Full of Obstacles}\label{subsec:office}

\begin{figure}
\centering
\includegraphics[width=3.5in]{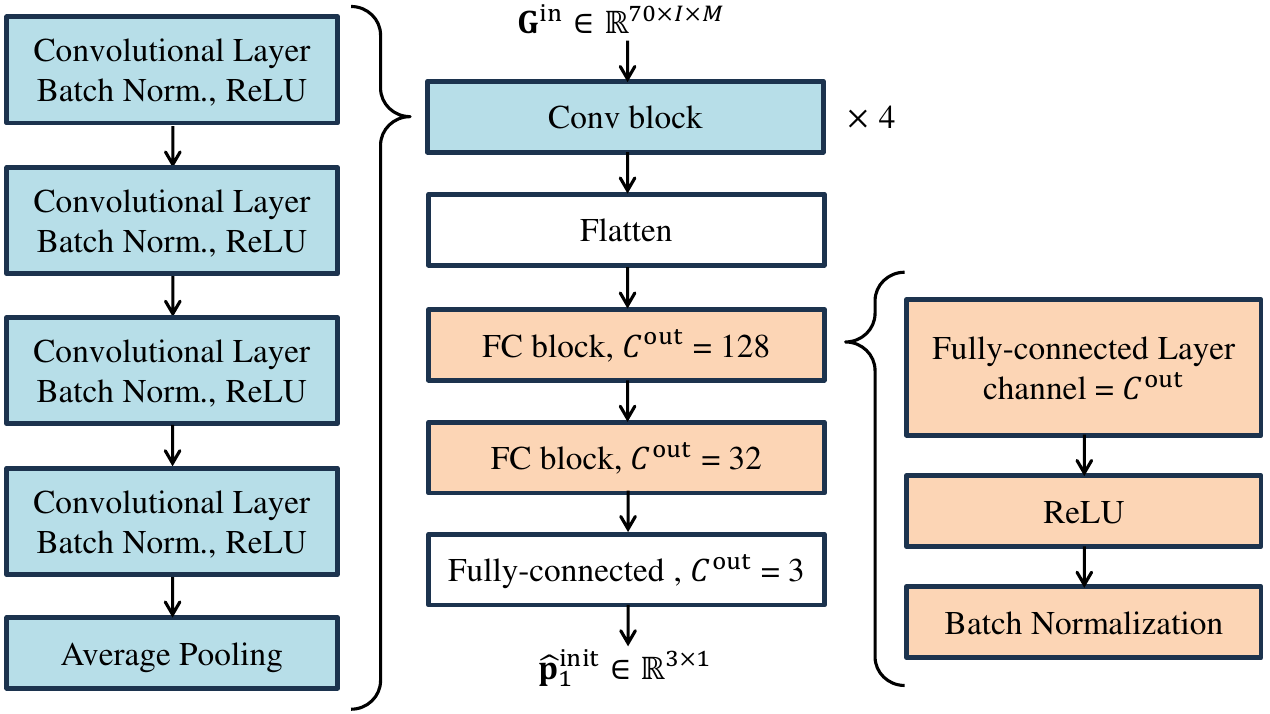}
\caption{A deeper CNN used in environment depicted in \cite[Fig. 9]{office}.}
\label{fig:ML_model_office}
\end{figure}

\begin{table}[!t]
\caption{Average localization error (m) of different localization methods tested in the environment depicted in \cite[Fig. 9]{office}. \label{tab:results_D}}
\centering
\begin{tabular}{ l c c c c c } \toprule
    & {\multirow{2}{*}{\makecell*[c]{Avg. run\\time (ms)}}}  &\multicolumn{4}{c}{Avg. localization error (m) }                                                             \\ \cmidrule(lr){3-6}
    &                                                                                           & IoT 10     & IoT 11      & IoT 12       & Total       \\ \midrule
                                               Two-step LS    & {\bf 2.2} & 0.88         & 3.05         & 4.08           & 2.67        \\ \midrule
      Two-step SLAM\cite{2step_EasyAPPos}& 81.4        & 0.95         & 2.24          & 3.45          & 2.21         \\ \midrule
                   OMP \cite{SOTA_RIS_NLoS}  & 9.5          & 0.85         & 2.69          & 2.71          & 2.08         \\ \midrule
          NOMP \cite{SOTA_ELAA2}, $L=1$ & 12.2        & 0.59         & 2.48          & 2.39          & 1.82         \\ \midrule
          NOMP \cite{SOTA_ELAA2}, $L=2$ & 18.7        & 0.55         & 2.43          & 2.29          & 1.76         \\ \midrule
          NOMP \cite{SOTA_ELAA2}, $L=3$ & 27.0        & {\bf 0.52}& 2.37          & 2.26          & {\bf 1.72} \\ \midrule
                    NM \cite{SOTA_NM}, $L=1$ & 55.0        & 0.54         & 2.57          & 2.46          & 1.86          \\ \midrule
                                  ML \cite{ML_RLA3} & 3.2          & 2.08         & 2.89          & 3.23          & 2.73          \\ \midrule
                                     ML-NOMP, $L=1$ & 6.5          & 1.83         & 2.04          & 2.4            & 2.09          \\ \midrule
                                     ML-NOMP, $L=2$ & 13.0        & 1.72         & 1.89          & 2.03          & 1.88          \\ \midrule
                                     ML-NOMP, $L=3$ & 21.3        & 1.63         & {\bf 1.81} & {\bf 1.98} & 1.81           \\ \midrule
                                          ML-NM, $L=1$ & 45.0        & 1.37         & 2.59          & 2.89          & 2.28           \\ \bottomrule

\end{tabular}
\end{table}

Although the room size in Fig. \ref{fig:Layout}(c) is larger than that of Fig. \ref{fig:Layout}(a) and \ref{fig:Layout}(b), it lacks obstacles or reflective surfaces other than the surrounding walls. To evaluate ML-NOMP in a more complex environment, we conduct simulations in an office filled with desks and walls, as depicted in \cite[Fig. 9]{office}, where IoT devices 10-12 are positioned at the locations of Tx 1-3 within the office. In our simulation, the UE does not move very far, making it challenging to locate distant devices. To address this, regardless of whether a wall blocks the signal between the UE and a device, we allow the UE to move within 10\,m of each IoT device and set $e^{\rm max} = 10\,{\rm m}$.

The obstacles in the office cause signal attenuation and complex reflections, reducing the localization accuracy of CNN in Fig. \ref{fig:ML_model} and making it inefficient to compensate for errors through refinement steps. Therefore, we use a deeper CNN with more convolutional layers, as shown in Fig. \ref{fig:ML_model_office}, to improve localization. The training data remains the same as in Section \ref{subsec:empty}. The convolutional layers and average pooling in this CNN model use the same kernel, channel, and stride as those in Fig. \ref{fig:ML_model}. As shown in Table \ref{tab:results_D}, this deeper CNN increases the running time by only 0.3\,ms compared to the previous model. Although the localization performance of the ML model alone is lower, ML-NOMP with just one iteration achieves the same performance as OMP. This observation is consistent with the results shown in Table \ref{tab:results_C}.

\begin{figure*}
\centering
\includegraphics[width=5.8in]{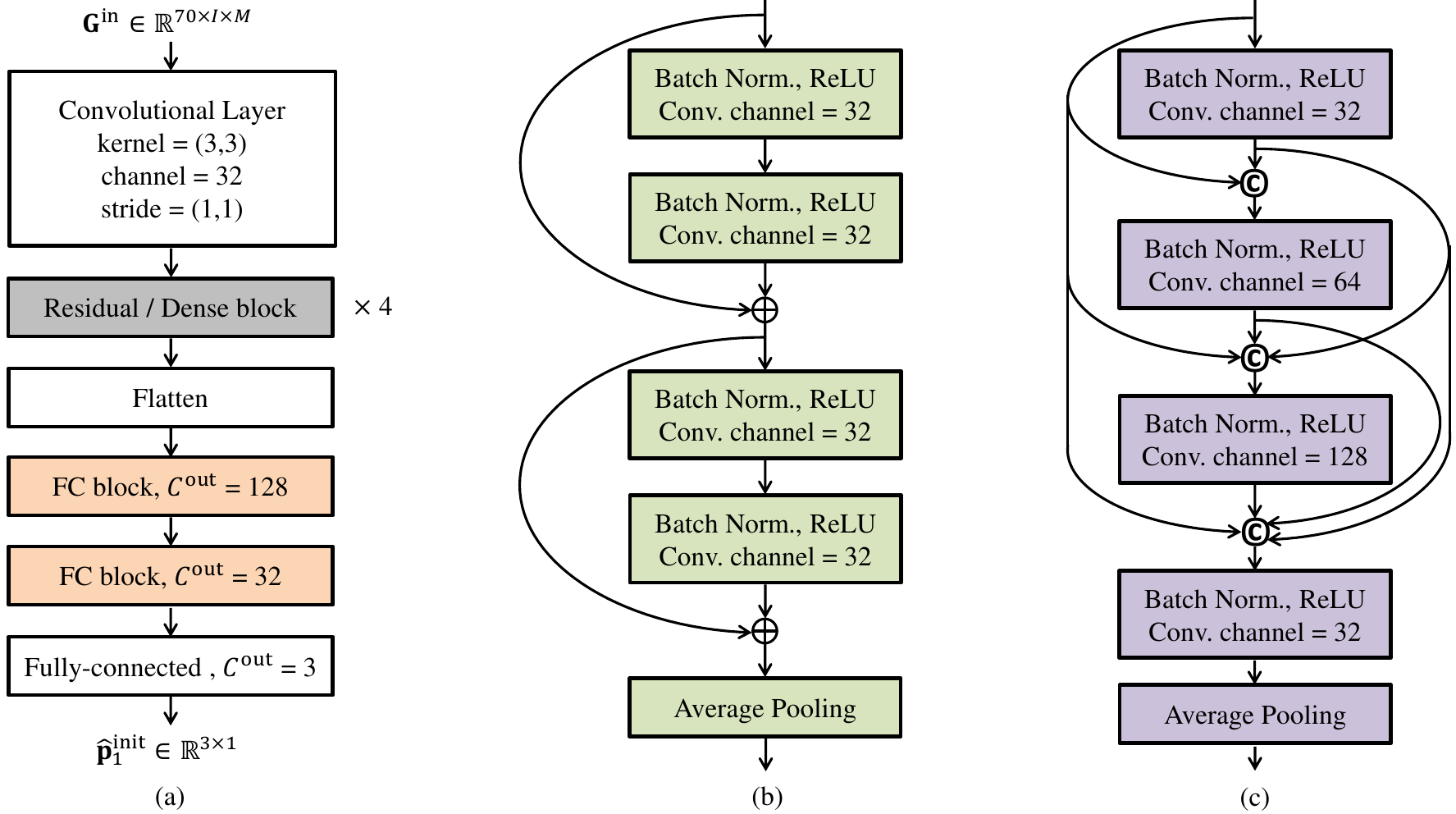}
\caption{(a) The primary architecture of ResNet and DenseNet. (b) A residual block in ResNet. (c) A dense block in DenseNet.}
\label{fig:Res&DenseNet}
\end{figure*}

\begin{table*}[!t]
\vspace{0.2cm}
\centering
\caption{Average running time (ms) and average localization error (m) for different ML models tested in the office environment shown in \cite[Fig. 9]{office}.  \label{tab:Diff_ML_time&Err}}
\begin{tabular}{ l c c c c c c} \toprule
                                       &  \multicolumn{3}{c}{Avg. run time (ms)}  &  \multicolumn{3}{c}{Avg. localization error (m) }  \\ \cmidrule(lr){2-4} \cmidrule(lr){5-7}
                                       & CNN (Fig. \ref{fig:ML_model_office})  & ResNet  & DenseNet  & CNN (Fig. \ref{fig:ML_model_office})  & ResNet  & DenseNet \\ \midrule
                                ML  & {\bf 3.2}                                                & 3.4        & 4.3            & 2.73   & 2.52   & 2.44          \\ \midrule
        ML-NOMP, $L=1$ & 6.5                                                         & 6.7        & 7.6            & 2.09   & 2.03   & 2.02          \\ \midrule
        ML-NOMP, $L=2$ & 13.0                                                       & 13.2      & 14.1          & 1.88   & 1.85   & 1.86     \\ \midrule
        ML-NOMP, $L=3$ & 21.3                                                       & 21.5      & 22.4          & 1.81   & {\bf 1.80} & {\bf 1.80}      \\ \bottomrule
\end{tabular}
\end{table*}

\subsection{Different ML Models}

The previous subsection demonstrates that deeper neural networks can be beneficial in complex environments. Next, we evaluate the impact of the proposed ML-NOMP using different ML architectures when the number of neural network layers is similar. The alternative ML architectures are illustrated in Fig. \ref{fig:Res&DenseNet}(a). Specifically, we replace the convolutional blocks in Fig. \ref{fig:ML_model_office} with residual and dense blocks, resulting in ResNet and DenseNet \cite{Res_DenseNet} architectures, respectively. Similar to the convolutional blocks in Fig. \ref{fig:ML_model_office}, the residual and dense blocks also consist of four convolutional layers; these layers have a kernel size of (3,3), a stride of (1,1), and varying numbers of channels due to the different architectures, as shown in Figs. \ref{fig:Res&DenseNet}(b) and \ref{fig:Res&DenseNet}(c). To integrate the input data with the residual and dense blocks, we add a single convolutional layer before these blocks, as depicted in Fig. \ref{fig:Res&DenseNet}(a).

The number of learnable parameters in these three architectures increases progressively from the CNN in Fig. \ref{fig:ML_model_office} to ResNet and then DenseNet. The running times for these architectures, presented in Table \ref{tab:Diff_ML_time&Err}, show the expected increase in running time as the complexity of the ML architecture increases. Table \ref{tab:Diff_ML_time&Err} also reports the corresponding localization performance. The results indicate that while DenseNet provides the best localization performance among the architectures, it also incurs the longest running time. Although the accuracy of ML is influenced by both the depth and complexity of the ML architecture, when the depth is kept constant, the localization accuracy converges to a similar value across different ML architectures after a few NOMP iterations. This highlights the effectiveness of combining data-driven ML with model-based NOMP, thereby reducing the reliance on complex ML models.

\section{Conclusion}

Given the impracticality of embedding an RLA into a smartphone for directly localizing IoT devices, we propose leveraging the smartphone's passive movement to create a VLA. Through CRLB analysis comparing VLAs and RLAs, we identify scenarios where VLAs can achieve localization accuracy comparable to RLAs, establishing a new benchmark in the field. Additionally, we introduce ML-NOMP, an innovative direct localization approach that combines ML with NOMP. This method eliminates the need for environmental data, such as room dimensions, providing a flexible solution adaptable to various environments. While ML's accuracy may vary across different settings, incorporating NR steps significantly enhances its performance. Extensive simulations demonstrate ML-NOMP's effectiveness in diverse indoor settings, highlighting its robustness, reduced computational load, and independence from prior environmental data, thereby paving the way for future advancements in IoT device localization.

This research focuses on scenarios where users unconsciously form the VLA through short-term movements, minimizing the impact of IMU error accumulation on localization accuracy, which can therefore be disregarded. The CRLB analysis indicates that localization accuracy improves as the UE moves farther; however, this also increases the accumulation of internal sensor errors.
In this context, ML can not only estimate the PS location but also aid in correcting internal sensor errors \cite{IMUfuture}. Future work will further integrate ML-based sensor correction methods with the proposed VLA localization technique.

\section*{Appendix A}

In \eqref{eq:VLA_single_FIM}, the partial derivative with respect to \eqref{eq:VLA_parameters} are:
    \begin{flalign}
    \stepcounter{equation}
    \label{eq:VH}
    \frac{\partial \left[ \qH_i \right]_{m,n}}{\partial \qp}               &= \frac{ {\sf j} 2 \pi}{\| \tilde\qp_i \|} \left( \frac{\tilde\qq_{i,m}}{\lambda} - \frac{\tilde\qp_i \tilde\qp_i^{\rm T} \tilde\qq_{i,m}}{\lambda \| \tilde\qp_i \|^2} - \frac{ f_n \tilde\qp_i}{c}   \right) \left[ \qH_i \right]_{m,n} &\notag, \\
    \frac{\partial \left[ \qH_i \right]_{m,n}}{\partial \tau_{i'}}        &= \begin{cases} {\sf j} 2 \pi f_n \left[ \qH_i \right]_{m,n},  & \mbox{if } i = i', \\
                                                                                                                                                               0,   & \mbox{if } i \neq i', \end{cases} &\notag\\
    \frac{\partial \left[ \qH_i \right]_{m,n}}{\partial \gamma_{i'}} &= \begin{cases} \frac{1}{\gamma_i} \left[ \qH_i \right]_{m,n}, & \mbox{if } i = i', \\
                                                                                                                                                               0,   & \mbox{if } i \neq i', \end{cases} &\tag{49}\\
    \frac{\partial \left[ \qH_i \right]_{m,n}}{\partial \beta_{i'}}      &= \begin{cases} {\sf j} \left[ \qH_i \right]_{m,n}, & \mbox{if } i = i', \\
                                                                                                                                                               0,   & \mbox{if } i \neq i', \end{cases} &\notag
    \end{flalign}
where $\tilde\qp_i=\qp-\qdelta_i$, and $\tilde\qq_{i,m}=\qOmega_i \qq_m$. Substitute \eqref{eq:VH} into \eqref{eq:VLA_single_FIM} can further obtain \eqref{eq:VJ_i}.

\begin{figure*}[!t]
    \begin{flalign}
    \stepcounter{equation}
    \label{eq:VJ_i}
    \qJ_i^{\qp,\qp}                     &=  \frac { 8 \pi^2 \gamma_i^2 } { \sigma^2 \| \tilde\qp_i \| ^2  }
                                                      \left(
                                                             \frac{N}{\lambda^2} \sum_{m=1}^{M} \left(
                                                                                                        \tilde\qq_{i,m} \tilde\qq_{i,m}^{\rm T}
                                                                                                        - \frac{ \tilde\qp_i^{\rm T} \tilde\qq_{i,m} \left( \tilde\qq_{i,m} \tilde\qp_i^{\rm T} + \tilde\qp_i \tilde\qq_{i,m}^{\rm T}\right)}{\| \tilde\qp_i \|^2}
                                                                                                        + \frac{(\tilde\qp_i^{\rm T} \tilde\qq_{i,m})^2 \tilde\qp_i \tilde\qp_i^{\rm T} }{\| \tilde\qp_i \|^4} \right)
                                                             + \frac{\Delta f^2 MN (N+1) (N+2) \tilde\qp_i \tilde\qp_i^{\rm T} }{12c^2}
                                                       \right),  &\notag \\
    \qJ_i^{\qp,\tau_{i'}}             &= \begin{cases} - \frac{2 \pi^2 \gamma_i^2 \Delta f^2 MN (N+1) (N+2) \tilde\qp_i  }{3c\sigma^2  \|\tilde\qp_i\| } , & \mbox{if } i = i', \\
                                                                           0,    & \mbox{if } i \neq i', \end{cases} &\notag\\
    \qJ_i^{\qp,\beta_{i'}}            &= \begin{cases} \frac { 4 \pi \gamma_i^2 N } { \sigma^2 \lambda \|\tilde\qp_i\|   }  \left( \sum_{m=1}^{M}\tilde\qq_{i,m}
                                                                           - \frac{  \tilde\qp_i  }{ \|\tilde\qp_i\|^2} \sum_{m=1}^{M} \tilde\qp_i^{\rm T}\tilde\qq_{i,m}  \right), & \mbox{if } i = i', \\
                                                                           0,    & \mbox{if } i, \neq i', \end{cases} &\notag\\
    \qJ_i^{\tau_{i'},\tau_{i''}}     &= \begin{cases} \frac{2 \pi^2 \gamma_i^2 \Delta f^2 MN (N+1) (N+2)}{ 3 \sigma^2}, & \mbox{if } i = i'  \mbox{ and } i = i'', \\
                                                                             0,    & \mbox{if } i \neq i' \mbox{ or } i \neq i'', \end{cases} &\tag{50} \\
    \qJ_i^{\beta_{i'},\beta_{i''}} &=  \begin{cases} \frac{2 \gamma_i^2 MN}{ \sigma^2}, & \mbox{if } i = i'  \mbox{ and } i = i'', \\
                                                                             0,    & \mbox{if } i \neq i' \mbox{ or } i \neq i'', \end{cases} &\notag\\
    \qJ_i^{\tau_{i'},\beta_{i''}}   &= 0, \forall i', \forall i'',  &\notag\\
    \qJ_i^{\gamma_{i'},\qvarepsilon_s}  &= 0, \forall i', \forall s.&\notag
    \end{flalign}
    \vspace*{4pt}
    \hrulefill
\end{figure*}

\section*{Appendix B}

In \eqref{eq:RLA_single_FIM}, the partial derivative with respect to \eqref{eq:RLA_parameters} are:
    \begin{flalign}
    \label{eq:RH}
    \frac{\partial \left[ \bar{\qH}_i \right]_{m,n}}{\partial \qp}               &= \frac{ {\sf j} 2 \pi}{\| \tilde\qp_i \|} \left\{ \frac{\tilde\qq_{i,m}}{\lambda} - \frac{\tilde\qp_i \tilde\qp_i^{\rm T} \tilde\qq_{i,m}}{\lambda \| \tilde\qp_i \|^2} - \frac{ f_n \tilde\qp_i }{c}   \right\} \left[ \bar{\qH}_i \right]_{m,n}, &\notag\\
    \frac{\partial \left[ \bar{\qH}_i \right]_{m,n}}{\partial \tau}              &= {\sf j} 2 \pi f_n \left[ \bar{\qH}_i \right]_{m,n}, &\notag\\
    \frac{\partial \left[ \bar{\qH}_i \right]_{m,n}}{\partial \gamma_{i'}} &= \begin{cases} \frac{1}{\gamma_i} \left[ \bar{\qH}_i \right]_{m,n}, & \mbox{if } i = i', \\
                                                                                                                                        0,   & \mbox{if } i \neq i', \end{cases} &\\
    \frac{\partial \left[ \bar{\qH}_i \right]_{m,n}}{\partial \beta_{i'}}     &= \begin{cases} {\sf j} \left[ \bar{\qH}_i \right]_{m,n}, & \mbox{if } i = i', \\
                                                                                                                                       0,   & \mbox{if } i \neq i'. \end{cases} &\notag
    \end{flalign}
Substitute \eqref{eq:RH} into \eqref{eq:RLA_single_FIM} can further obtain
    \begin{align}
    \label{eq:RJ_i}
    \bar{\qJ}_i^{\qp,\qp}                       &= \qJ_i^{\qp,\qp},   \notag\\
    \bar{\qJ}_i^{\qp,\tau}                      &= \qJ_i^{\qp,\tau_i}, \notag \\
    \bar{\qJ}_i^{\qp,\beta_{i'}}             &= \begin{cases} \qJ_i^{\qp,\beta_i}, & \mbox{if } i = i', \\
                                                                                        0,   & \mbox{if } i \neq i', \end{cases} \notag\\
    \bar{\qJ}_i^{\tau,\tau}                     &= \qJ_i^{\tau_i,\tau_i},  \\
    \bar{\qJ}_i^{\beta_{i'},\beta_{i''}}  &= \begin{cases} \qJ_i^{\beta_i,\beta_i}, & \mbox{if } i = i'  \mbox{ and } i = i'', \\
                                                                                      0,    & \mbox{if } i \neq i' \mbox{ or } i \neq i'', \end{cases} \notag\\
    \bar{\qJ}_i^{\tau,\beta_{i'}}                    &=  0,  \forall i', \notag\\
    \bar{\qJ}_i^{\gamma_{i'},\qvarepsilon_s} &= 0, \forall i', \forall s. \notag
   \end{align}

{\renewcommand{\baselinestretch}{1.06}
\bibliographystyle{IEEEtran}
\bibliography{IEEEabrv,References}

\end{document}